\documentclass[sn-mathphys]{sn-jnl}

\newcommand{\of}{\textsc{OpenFOAM}\textsuperscript\textregistered}
\newcommand*{\bs}[1]{\ifmmode\boldsymbol{#1}\else\textbf{#1}\fi}

\begin{document}

\title[Pescimoro et al.]{Emergence of non-Fickian transport in truncated pluri-Gaussian 
permeability fields}
\author*[1]{\fnm{Eugenio} \sur{Pescimoro}}\email{eugenio.pescimoro@gmail.com}
\author[1]{\fnm{Matteo} \sur{Icardi}}\email{matteo.icardi@nottingham.ac.uk}
\author[2]{\fnm{Giovanni} \sur{Porta}}\email{giovanni.porta@polimi.it}
\author[3]{\fnm{Marco} \sur{Bianchi}}\email{marcob@bgs.ac.uk}
\affil*[1]{\orgdiv{Industrial and Applied Mathematics}, \orgname{University of Nottingham}, 
\orgaddress{\street{University Park}, \city{Nottingham}, \postcode{NG7 2RD}, 
\state{Nottinghamshire} \country{United Kingdom}}}
\affil[2]{\orgdiv{Civil and Environmental Engineering}, \orgname{Politecnico di Milano}, 
\orgaddress{\street{Piazza L. Da Vinci, 32}, \city{Milan}, \postcode{20133}, \state{Lombardia}, 
\country{Italy}}}
\affil[3]{\orgname{British Geological Survey}, \orgaddress{\street{Nicker Hill}, \city{Keyworth}, 
\postcode{NG12 5GG}, \state{Nottinghamshire}, \country{United Kingdom}}}


\abstract{We present a numerical simulation study of advective-diffusive scalar transport in 
three-dimensional high-contrast discontinuous permeability fields, generated with a 
truncated pluri-Gaussian geostatistical approach. A range of permeability contrasts, correlation 
lengths, and Péclet numbers are studied to characterise the transition to non-Fickian transport 
behaviour.
This is triggered by high permeability contrasts between different zones and is enhanced by the presence of connected higher permeability channels, which are characterised by high 
advective flow velocities. In this case, the overall transport behaviour deviate from the macroscopic advection-dispersion model based on a Fickian analogy.  The numerical experiments are run with an Eulerian approach using a novel unified numerical 
framework based on the finite-volume library \of, for i) generating random pluri-Gaussian porous 
media, ii) solving the steady state Darcy-scale flow, iii) solving the advection diffusion 
equation, iv) computing post-processing quantities such as first order statistics, spatial 
probability density functions and breakthrough curves. We identify a hierarchy of non-Fickian transport triggering factors, the strength of permebility contrast being the pivotal driver. The Péclet number and the characteristic length at which facies transitions are observed as secondary factors. Transport remains Fickian when the facies conductivities differ by up to one order of magnitude. Greater permeability contrasts act strengthen the emergence of fast flow channels leading to non-Fickian transport.}


\maketitle

					                                                            

\section{Introduction}\label{sec1}

Subsurface flow and solute transport modelling is used in several engineering and environmental fields 
(CO\textsubscript{2} storage, groundwater remediation, oil recovery) where mathematical and computational models 
play a central role in supporting the reliability of analysis and design strategies. The 
effectiveness of advection-dispersion models in describing solute transport in highly heterogeneous 
media such as geological formations has been questioned
\cite{adams1992field, barlebo2004investigating, fiori2016debates}, and the definition of appropriate 
models and their parameterization remains an open field of research
\cite{zinn2003good, jankovic2017good, bianchi2016lithofacies}. An important challenge is how to simulate
 non-Fickian behaviour, which originates mainly from physical 
heterogeneities emerging across multiple scales 
\cite{dentz2011mixing, gelhar1983three}. Transport is defined as anomalous or non-Fickian when 
solute plumes and breakthrough curves display a significant departure from the predictions made by 
an advective-dispersive model where dispersion is expressed with a Fickian analogy, i.e. is lumped 
with molecular diffusion in a single effective coefficient 
\cite{bear2012hydraulics}.

Approaches to modelling solute transport in heterogeneous porous media largely differ depending on 
the scale of interest. In this work we start from a mesoscale, which correspondins to a resolution 
where geological media can be described by an equivalent continuum with spatially heterogeneous 
properties 
\cite{DeBarros21features,Riva2008JCH}. At this scale, solute transport is governed by two separate 
mechanisms: advection and local hydraulic dispersion which includes the contributions of molecular 
diffusion and mechanical dispersion. At the mesoscale, spatial heterogeneity is explicitly 
modelled, most commonly using a statistical characterization.
We then move to macroscale modelling, where the aim is to define an effective model able to 
describe the dynamics of the system without an explicit description of the underlying 
heterogeneity. In classical descriptions 
\cite{dagan_book}, velocity at these scales may be interpreted as the average Darcy velocity while 
the hydraulic dispersion coefficient turns into a macrodispersion coefficient, employed to quantify 
the effect of heterogeneity on solute spreading. This model has been questioned in the literature 
and alternative non-Fickian effective models have been proposed 
\cite{hansen2018direct, zech2021comparison, neuman2009perspective}. These approaches mainly focused 
on cases where the underlying (mesoscale) log-conductivity field has a Gaussian distribution. Beyond this specific 
case, the validity of the Advection Dispersion Equation (ADE) based macrodispersive models 
are not clearly identifiable a priori, although they are certainly heavily controlled by the degree of 
heterogeneity of porous media properties and their spatial organisation
\cite{neuman2009perspective}. Discontinuous permeability fields with a high connectivity degree and 
sharp contrast between regions are recognised among the most important factors that regulate the 
transport of solute
\cite{zhang2013impact}. Aquifers characterized by highly connected \textit{facies}, i.e. finite 
portion of the subsurface with similar physical properties, and high permeability contrasts are 
characterised by fast flow channels with persistent spatial velocity correlations, which make the overall solute 
behaviour non-Fickian
\cite{bianchi2016lithofacies}. The persistence of this anomalous transport behaviour at the macroscale can 
be due, for example, to these regions where the flow paths create preferential fast channels 
\cite{edery2014origins}, a feature that also influences reactive transport settings 
\cite{edery2016chara,edery2021feed}.

In this work we investigate solute transport and the onset of anomalous or non-Fickian transport 
behaviour in high-contrast heterogeneous permeability fields, generated with the geostatistical 
pluri-Gaussian truncated (PGS) method 
\cite{mariethoz2009truncated}. Solute transport has been widely investigated in continuous Gaussian 
and non-Gaussian permeability fields 
\cite{Gotavac2009,SoleMari2021Solute}, and methods have also been proposed to handle non-continuous 
fields, suitable to reproduce geomaterials where property transition is marked by sharp interfaces 
\cite{bianchi2017geological}. PGS random fields are used in this context to model actual subsurface geological 
media in a sedimentary setting. In this context 
this model is used to link an assumed geological architecture or structure, e.g. driven by sedimentological 
rules, with the spatial distribution of physical properties such as porosity or hydraulic conductivity.
This allows to create fields starting from given geological assumptions and explicitly control the 
connectivity of high- and low-permeability facies. Therefore, PGS can be 
employed to reproduce and interpret the emergence of non-Fickian transport traits observed in real 
geological media. However, to our knowledge, studies that systematically address the impact of PGS field 
paramterization with solute transport features are lacking. Our aim here is to fill this 
gap and investigate the connectivity and permeability contrast thresholds that drive a transition 
between Fickian and non-Fickian. We quantify the 
deviation of the results obtained from numerical simulations in PGS domains from Fickian behavior 
by comparing them to the analytical solution of the advection-dispersion equation. This allows us 
to identify the physical and structural characteristics of the geological media that can lead to 
non Fickian transport and ultimately paves the way 
towards aquifer typing approaches where non Fickian transport features may become identifiable from 
a knowledge of the properties of the field.

To achieve these objectives we rely on numerical transport simulations, by solving the 
advection-dispersion equation in heterogeneus media using an Eulerian finite volume method. This 
approach is implemented as a parallel open-source code based on \of, as part of the SECUReFoam 
library 
\cite{SECUReFoam}. The advantages of the Eulerian approach are that it allows the 
computation of Péclet number and that an accurate simulation of solute low concentration tails does not require a large 
particle ensemble as with Lagrangian formulations, which have often been used in the recent 
literature 
\cite{edery2014origins, savoy2017geological, dentz2011mixing}. Morever, the Eulerian description is 
closer to the experimental conditions where results are often obtained in terms of molar or mass 
concentration while Lagrangian approaches need to be postprocessed to obtain local concentration 
fields.

From an operational perspective, our approach is based on a single computational framework, 
including a geostatistical algorithm for permeability field generation, a numerical code for flow 
and transport simulation, and 
post-processing tools.  This is an interesting feature of our approach as the synthetic generation 
of realistic geological domains remains one of the main challenges in modelling flow and transport
\cite{hesse2014generating}. Several approaches are available to reproduce complex subsurface 
structures (sequential Gaussian simulations \cite{dimitrakopoulos2004generalized}, Markov chain 
probability
\cite{carle1997modeling}, Multiple-point statistics 
\cite{strebelle2002conditional}) as well as a number of geostatistical open toolboxes (GSLib
\cite{deutsch1998gslib}, T-PROGS
\cite{carle1999t}). Nevertheless, few tools exist that provide integrated geostatistical, flow 
and transport simulation solvers (OpenGeoSys
\cite{kolditz2012opengeosys}, porousMultiphaseFoam
\cite{horgue2015open}, DuMux
\cite{flemisch2011dumux}).


This work is structured as follows: in Section 2 we give the mathematical overview of the problem, 
in Section 3 we describe the testcases and summarise the numerical methodologies. Numerical results 
and the post-processing are presented in Section 4, before we draw conclusions and give some 
guidelines about the emergence of non-Fickian transport. For the sake of clarity, the terms 
``facies'' (uncountable) and ``category'' as well as ``lithotype'' and ``truncation'' rule will be 
used interchangeably depending on the context.



\section{Methods}
We describe here the methods underpinning our numerical simulations. We start by presenting the 
geostatistical framework and then move to the description of the physical problem, i.e. the flow 
and transport setting.

\subsection{Geostatistical model}
The permeability fields used in this work are generated via the pluri-Gaussian Simulation 
(PGS) method, i.e. applying a truncation rule to continuous multivariate Gaussian random fields 
(GRF) 
\cite{mariethoz2009truncated}. Fields generated with this approach are characterised by:
\begin{itemize}
	\item discontinuous permeability fields characterised by a discrete number of zones of 
uniform permeability whose spatial arrangement is the result of a specific truncation rule (i.e. 
Lithotype rule \cite{armstrong2011plurigaussian});
	\item high geological realism since the truncation rule allows to simulate observed  geometrical relations between geological facies \cite{koltermann1996heterogeneity, linde2015geological, armstrong2011plurigaussian}.
\end{itemize}

GRFs can be generated in the frequency domain by multiplying independent complex 
Gaussian random variables by the spectral representation of the covariance function. The spatial 
field is then reconstructed by applying the inverse Fourier transform to the spectral GRF. To 
ensure independence of the random field generation from the mesh-discretisation and to allow 
arbitrary unstructured grids, we apply an explicit discrete inverse Fourier transform 
discretised with $N_f$ frequencies in each direction instead of the IFFT algorithm. Following 
\cite{mandelbrot1968fractional, hesse2014generating}, a discrete-in-frequencies, 
continuous-in-space representation of a complex GRF is therefore given by:
\begin{equation} \label{eq:DiscreteFT}
	Z(\bs{x}) = \sum_{j=0}^{N_f} \cos(2 \pi \bs{a}_j\cdot \bs{x}) \sqrt{S(\bs{a}_j)} W_j + i 
\sum_{j=0}^{N_f} \sin(2 \pi \bs{a}_j \cdot\bs{x}) \sqrt{S(\bs{a}_j)} W'_j
\end{equation}
where $\bs{x}$ is the position vector, $\bs{a}_j=(a_{x,j},a_{y,j},a_{z,j})$ is the $j^{th}$ 
frequency vector, $W_j$ and $W'_j$ are independent complex Gaussian random variable and 
$S(\bs{a}_j)$ is the amplitude of the spectral measure.
From $Z$, we can then extract two independent Gaussian random fields from its real and imaginary 
parts and compute it on an arbitrary spatial discretisation.


The covariance function of a stationary field quantifies the covariance  $\gamma(\bs{r})$ between a 
pair of values of a random variable located at points separated by the distance $\bs{r}$. The correlation function 
$\rho(\bs{r})$ is the covariance function scaled by the variance $\sigma^2$, i.e., $\gamma(\bs{r}) 
= \sigma^2 \rho(\bs{r})$.

In this work, we assume an exponential correlation function
\begin{equation} \label{eq:expSpe}
	\rho(\bs{r}) = 1 - e^{
	-\sqrt{\frac{r_x^2}{\lambda_x^2} + \frac{r_y^2}{\lambda_y^2}
	+ \frac{r_z^2}{\lambda_z^2}}
	}
	\end{equation}
	and corresponding spectrum
	\begin{equation}
	S(\bs{a}) = \sigma^2 \|\bs{\lambda}\|^d \frac{\Gamma \left( \frac{d+1}{2} \right)}{\left( 
\pi \left(1 + a_x^2\lambda_x^2
	+ a_y^2\lambda_y^2 + a_z^2\lambda_z^2\right) \right)^{\frac{d+1}{2}}}\,,
\end{equation}
where $d=3$ is the number of dimensions, $\Gamma$ is the Gamma function,  
$\bs{\lambda}=(\lambda_x,\lambda_y,\lambda_z)$ are the correlation lengths.

\begin{figure}[!htbp]
	\centering
	\includegraphics[width=1.0\linewidth]{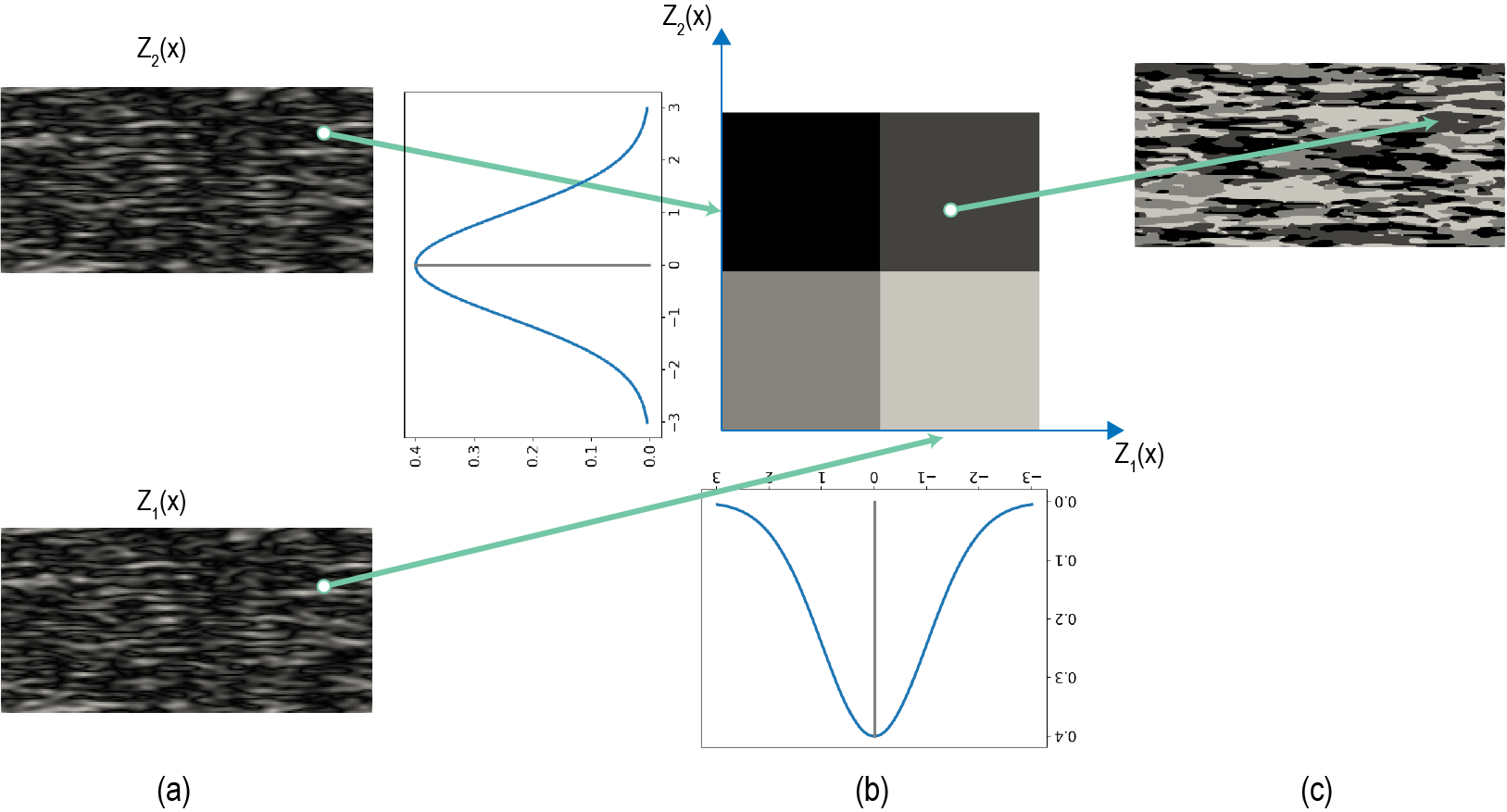}
	\caption{Truncated pluri-Gaussian simulation. a) Continuous multivariate Gaussian random 
fields $Z_1$ and $Z_2$ generation; b) truncation rule for four facies domain and its corresponding 
thresholds on the Gaussian distribution of the variables; c) sample of a two-dimensional truncated 
pluri-Gaussian random field. The arrows indicate the contribution of the two GRFs in assigning a 
given category at a selected location in space.}
	\label{fig:PGSmethod}
\end{figure}

GRFs are continuous fields, but geological media are often characterised by abrupt changes in 
physical and chemical properties. With the PGS approach discontinuous patterns are reproduced from
the truncation of two GRFs according to a \textit{lithotype} or \textit{truncation rule} (fig. 
\ref{fig:PGSmethod}), which bins continuous values into a set of categories.
The smooth transition which characterizes the GRF is then replaced by $n = (N_r+1)(N_s+1)$ 
categorical values where $N_r$ and $N_s$ are the number of thresholds applied via the truncation 
rule to the two GRFs. In this sense, the ``truncated'' adjective refers to a GRF that has been 
discretized through a binning process. The probability, i.e. the proportion, of the facies 
$\varphi_{i}$ is obtained from
\begin{equation}
	p_{\varphi_i}(\bs{x}) = \left[ G(r_i) - G(r_{i-1}) \right] \left[ G(s_i) - G(s_{i-1}) 
\right] \quad i=1 \ldots n
\end{equation}
where $n$ is the number of categories, $G$ is the cumulative Gaussian distribution with the mean 
and the variance typical of each field. The lithotype rule allows to control the chances of a certain category 
of being into contact with the others. This constitutes a fundamental feature as it allows the simulated field 
to reflect geological transition patterns observed in field data. Typically, transition patterns are 
captured along the vertical direction by processing field sample information through transition 
probability matrices \cite{carle1996transition, weissmann1999three} while field observations and/or established conceptual models of geological environments are used as guidance for the estimation of transition patterns in the horizontal directions \cite{armstrong2011plurigaussian}.
In this work, we assume the single truncation diagram, in Fig.~\ref{fig:PGSmethod}, 
while changing the correlation lengths $\bs{\lambda}$ of the underlying GRFs and the permeability values 
assigned to different categories. These four categories have also equal probability and therefore volumetric fractions 
$p_{\varphi_i} = 25 \%$. The distribution of the multivariate random variables adopted to generate the underlying continuous 
Gaussian random fields in this study has mean $\mu=0$ and $\sigma=1$ and their correlation function is exponential with 
different values of correlation length.

\subsection{The flow model}
We assume fluid flow obeys the standard Darcy's equation which reads
\begin{equation} \label{eq:Darcy}
	\bs{V} = - \frac{\bs{k}}{\mu} (\nabla p + \rho g \nabla z),
\end{equation} 
where $\bs{V}$ is the Darcy velocity vector $[LT^{-1}]$, $\bs{k}$ is the permeability tensor 
$[L^2]$, $\mu$ is the dynamic viscosity $[ML^{-1}T^{-1}]$, $p$ is the pressure $[M T^{-2} L^{-1}]$, 
$\rho$ is the fluid density $[M L^{-3}]$, $g$ is the gravity constant $[LT^{-2}]$ and $\nabla z = 
(0, 0, 1)$ $[-]$ is an upward unit vector. For this study we set $g=0$ as any influence of the 
solute on the liquid density is assumed to be negligible.


The flow solver implemented in \of is based on equation \eqref{eq:Darcy} assuming an incompressible 
fluid. Therefore pressure can be computed according to a Poisson equation 
\begin{equation} \label{eq:poisson}
	\nabla\cdot\bs{V} = - \nabla \cdot \frac{\bs{k}}{\mu} \nabla p = 0
\end{equation} 
where we have assumed no sources or sinks are present and the gravity term is zero. The 
permeability tensor is, from this point, assumed diagonal and isotropic, i.e., 
$\bs{k}=k\mathbb{I}$, $\mathbb{I}$ being the identity matrix.


\subsection{Local transport model}
The advective flux per unit area $\bs{J}_{adv}$ $[L T^{-1}]$ is the product of the advective Darcy 
velocity $\bs{V}$ $[L T^{-1}]$ and solute concentration $c \ [-]$
\begin{equation}
	\bs{J}_{adv} = \bs{V} c.
\end{equation}\par
In line with previous work \cite{edery2014origins}, we neglect mechanical dispersion and model the 
diffusive fluxes $\bs{J}_{mol}$ $[L T^{-1}]$ as
\begin{equation}
	\bs{J}_{mol} = -\phi \bs{D}_{mol} \nabla c
\end{equation}
where $\bs{D}_{mol}$ $[L^2 T^{-1}]$ is the molecular diffusion tensor and $\phi$ is the porosity of 
the medium.
Summing up the advective and diffusive fluxes, the conservation of mass
yields the advection-diffusion equation, which, for the case of isotropic diffusion and porosity 
and no source/sink terms is
\begin{equation} \label{eq:ADiff}
	\frac{\partial c}{\partial t} + \nabla \cdot (\bs{v} c) - \bs{D}_{mol} \nabla^2 c = 0
\end{equation}
where $\bs{v} \equiv \bs{V}/\phi$ is the fluid velocity, i.e. the velocity that would be measured 
by a flow meter in the porous domain and $\bs{D}_{mol} = D \mathbb{I}$. The concentration boundary 
conditions adopted to solve (\ref{eq:ADiff}) are fixed single concentration on the inlet, no flux 
on all lateral sides and zero gradient on the outlet of the domain. In this study, to focus on the 
effects of the heterogeneity, the geostatistical parameters and the Péclet number, we have made 
strong assumptions on the permeability (isotropic and diagonal), porosity (constant) and neglected 
mesoscopic dispersion. Whilst preliminary tests suggested these do not impact the main findings of 
this work, the investigation of these processes may be tackled in future contributions.
\subsection{Macrodispersion model}
Transport mechanisms described so far characterise the transport behaviour at 
the mesoscale, i.e. where geological and flow resolution allows for heterogeneity to be modelled 
explicitly. However, macroscale models aim to provide an overall description while using an 
effective/upscaled advection-dispersion equation neglecting heterogeneity. Here, we only focus on 
transport along the main velocity direction and the longitudinal dispersion processes, therefore we 
will compare our results with a one-dimensional advection-dispersion equation:
\begin{equation} \label{eq:ADmacro}
	\frac{\partial C}{\partial t} + \overline{v_x} \dfrac{\partial C}{\partial x} - 
D^{mac}_{xx} \dfrac{\partial^2 C}{\partial x^2} = 0,
\end{equation}
where $C$ is the section-averaged concentration and the longitudinal component of the 
macrodispersion tensor is $D^{mac}_{xx}$. The $\overline{X}$ notation indicates the spatial average 
of the field $X$ at a given section, also known as section average.  Macrodispersion in Fickian 
transport models is often represented as the product between a typical length scale and an average 
velocity (\ref{eq:ExpDmac}).


\subsection{Quantities of interest}
The record of the section-averaged concentration in time at a control section (e.g. outlet boundary 
or an arbitrary point) constitutes the breakthrough curve (BTC). Under a continuous injection, the 
BTC is equivalent to the cumulative density function (CDF) of the arrival times of the solute mass ($F(t)$) while its 
time derivative, which is a concentration rate, is the probability density function (PDF) of the 
arrival times ($f(t)$). These functions are typically obtained by injecting a pulse in time or a constant 
concentration at the inlet (or an injection point). Due to the linearity of the advection-diffusion 
equation, the system is fully characterised by the solution in time of a single injection (i.e. its 
Green's function).

To enable the comparison between simulations considering different parameters and different 
durations, we consider a dimensionless time $T$ obtained by dividing $t$ by the average travel time, 
calculated as the ratio between the longitudinal domain dimension and the average fluid velocity. This 
quantity is equivalent to the injected pore volume. The section averaged concentration at the 
outlet is non-dimensionalised by dividing it by the fixed single inlet concentration and is 
represented by $\overline{c}$. 

In the post processing phase of the simulation results, the following quantities were estimated:
\paragraph{Péclet number}\par
\begin{equation} \label{eq:Pe}
	Pe_x \ [-] \ = \dfrac{\overline{v_x} \lambda_x}{D_{mol}};
\end{equation} 
\paragraph{effective permeability}\par
\begin{equation} \label{eq:kEff}
    {k}^{eff}_{x} \ [m^2] \ = - \frac{\overline{v_x} \mu}{\frac{\partial p}{\partial x} - \rho g};
\end{equation}
\paragraph{nominal macrodispersion}\par
\begin{equation} \label{eq:ExpDmac}
	D^{ij}_{mac} \ [m^2/s] \ = \phi \bs{\lambda}^T \bs{V}
\end{equation}
where $\bs{\lambda}$ and $\bs{V}$ are typical lengths and velocity vectors. Equation 
(\ref{eq:ExpDmac}) allows the macrodispersion matrix to be approximated a priori starting from 
geostatistical (correlation length $\bs{\lambda}$) and flow (velocity $\bs{V}$) data, independent 
of the BTC data. Concentration data coming from the BTC constitutes the basis for the methods adopted 
to estimate the macrodispersion a posteriori, as illustrated in section \ref{sec:BTC_InvGau}.
  
\subsubsection{Breakthrough curve and inverse Gaussian approximation} \label{sec:BTC_InvGau}
The mass arrival time distribution simulated with the one-dimensional advection-dispersion equation is 
the inverse Gaussian distribution. This corresponds to the analytical solution of eq. 
\eqref{eq:ADmacro} in a semi-infinite one-dimensional domain with a Dirac-delta initial condition. 
For practically relevant parameters, this is almost indistinguishable from the solution on a finite 
domain with a Dirac-delta (in time) concentration injection at the inlet. For our problem with a 
continuous injection at the inlet, due to the linearity of the problem, the BTC is well 
approximated by the integral in time of the Inverse Gaussian distribution, computed for a fixed 
section in space (the outlet in our case). When transport behaviour is Fickian, we can approximate 
the experimental BTCs with the cumulative density function of the Inverse Gaussian distribution as
\begin{equation}\label{eq:CumInvG}
    F(T; \mu_1, \nu) = \bar{c} = \Phi \left( \sqrt{\frac{\nu}{T}} \left( \frac{T}{\mu_1}-1 \right) 
\right) + \mathrm{e}^{\frac{2 \nu}{\mu_1}} \Phi \left( -\sqrt{\frac{\nu}{T}} \left( 
\frac{T}{\mu_1}+1 \right) \right)
\end{equation}
where $\Phi$ is the standard normal cumulative distribution function, $\mu_1$ is the first order 
statistical moment of the concentration rate distribution and $\nu$ is a shape parameter. The PDF 
of the solute arrival times can be obtained through a time derivative of \eqref{eq:CumInvG} and 
corresponds to the PDF of the solute arrival times. This PDF is expressed as 
\cite{tartakovsky2019diffusion}
\begin{equation} \label{eq:InvGau}
	f(T; \mu_1, \nu) = \frac{\partial \bar{c}}{\partial T} = \sqrt{\frac{\nu}{2 \pi T^3} exp 
\left[ -\frac{\nu(T-\mu_1)^2}{2 \mu_1^2 T} \right]}.
\end{equation}
Other analytical solutions are available for different boundary conditions on finite domains 
\cite{van1982analytical}. For the purposes of this paper, we will only consider the Inverse 
Gaussian model as a reference for Fickian transport due to its simpler analytical formula more 
suitable to fitting and moment matching. The inverse Gaussian is generally a good approximation
\cite{berkowitz2006modeling}
for macro-dispersion in heterogeneous media if
\begin{itemize}
	\item domain is large;
	\item experiment time is long;
	\item domain's properties are ergodic.
\end{itemize}
The PDF of mass arrival times from numerical experiments that follow a Fickian model are clearly 
distinguishable by a short and exponential-like tail as $t \rightarrow \infty$.
Non-Fickian transport processes have a clear impact on the shape of the PDF of the arrival times: 
early arrival concentrations raise the PDF peak and power low scaling emerges prior to exponential 
decay \cite{berkowitz2006modeling, edery2014origins}. 
of the domain, the last part of the curve will always follow an exponential-like trend because it 
represents the filling of the slowest remaining regions by pure diffusion.

\paragraph{The moments' method} \label{sec:MomMet}
Moments of the numerically computed BTC can be matched with the ones from the cumulative Inverse 
Gaussian distribution. Following 
\cite{yu1999moment, kreft1978physical}, the estimation of the statistical moments of the cumulative 
Inverse Gaussian is performed by approximating its parameters $\mu_1$ and $\nu$ as
\begin{align}
	E[\bar{c}] &= \mu_1 \\
	Var[\bar{c}] &= \mu_2 - \mu_1^2 = \frac{\mu_1^3}{\nu}.
\end{align}
To compute the first and second order moments we used
\begin{align}
	\mu_1 &= \int_0^{+\infty} f T dT = \int_0^{+\infty} F' T dT = -\int_0^{+\infty} F dT + 
\left[ FT \right]_0^{+\infty} \nonumber \\
	      &= - \sum_{i=0}^{+\infty} F_i \Delta T + F_{+\infty}T_{+\infty}, \\
	\mu_2 &= \int_0^{+\infty} f T^2 dT = \int_0^{+\infty} F' T^2 dT = -2 \int_0^{+\infty} F T 
dT + \left[ FT^2 \right]_0^{+\infty} \nonumber \\
	      &= -2 \sum_{i=0}^{+\infty} F_i T_i \Delta T + F_{+\infty}T_{+\infty}^2.
\end{align}
The estimated effective velocity and macrodispersion coefficient can be estimated from statistical 
moments as
\begin{align}
	V_x &= \frac{L_x}{\mu_1} \\
	D^{mac}_{xx} &= \frac{\mu_2 V^3}{2 L_x} \label{eq:Dmac}
\end{align}
where $L_x$ is the distance between the inlet and outlet sections (in our case the domain length). 
To quantify the distance between the numerical outputs and the Inverse Gaussian approximation, a 
normalised error $e$ was defined as
\begin{equation} \label{eq:RelErr}
	e = \dfrac{||\bar{c}(T) -F(T)||}{||\bar{c}(T)||} \cdot 100.
\end{equation}

\paragraph{Least squares estimation}
Parameter estimation is performed by minimising the least squared error between numerical data 
and the models \eqref{eq:CumInvG}-\eqref{eq:InvGau}. Under the assumption of identically 
distributed and uncorrelated errors this procedure corresponds to a maximum likelihood estimation. 
This procedure is applied to three types of data
\begin{itemize}
	\item probability density function of the concentration distribution, obtained by numerical 
differentiation of the BTC values at the outlet;
	\item cumulative density function of the concentration arrival times (i.e. the BTC);
	\item small subset of concentration data which spans 0.5 dimensionless time unit and it is 
centered around the peak of the probability density function.
\end{itemize}
In the first and third case, the analytical function used as reference to perform the least squares 
fitting is the probability density function of the inverse Gaussian distribution given by equation 
\eqref{eq:InvGau} while for the second case the analytical function is equation \eqref{eq:CumInvG}. 
For all cases the analysis was performed using Python library \texttt{lmfit} constraining 
the estimation so that $\mu_1$ and $\nu$ were always non-negative. Values of the estimated 
parameters uncertainty are also obtained from the diagonal entries of the parameters covariance 
matrix computed by \texttt{lmfit} and were used to assess the reliability of the estimate. The 
initial values for the least square estimation were set equals to values computed for $\mu_1$ and 
$\nu$ with the moments' method. 



\section{Numerical experiments} \label{sec:numexp}

Geostatistical, flow and transport numerical simulations were conducted over hexahedral domains 
which represent a portion of the subsurface with dimensions $(L_x/l, \ L_y/l, \ L_z/l) \ = \ (2, \ 
1, \ 1)$ where $L_i$ are the dimensions of the domain and we took $l = L_y = L_z$ as the reference 
length. The mesh is unstructured and characterised by cubic cells of dimension $d/l \ = \ 0.05$, so 
that the total number of cells is $2 \cdot 10^6$.

The permeability distribution within the domain corresponds to the field generated with a PGS 
simulation while porosity is assumed homogeneous over the 
domain. All the simulated fields considered in this study share the correlation function reported in 
equation \eqref{eq:expSpe}, the number of permeability zones, as well as the volumetric proportion 
for each of the facies (see table \ref{tab:constParam}). We investigate the variability of the 
observed output and of the estimated parameters as a function of three inputs: geostatistical 
parameters (e.g., correlation length used to generate the conductivity fields), hydraulic properties 
(i.e., permeability) and transport regime, defined in terms of $Pe$.

Based on the assigned permeability values, we distinguish two cases: low and high contrast. For both 
cases the permeability values $k_i$ assigned to the four geomaterials considered are evenly spaced 
on a logarithmic scale. However, for the low permeability contrast case the four permeability 
values range between $10^{-10}$ and $10^{-13} \ m^2$ with a relative ratio $log_{10}(k_i/{k_{i+1}}) 
= 1$ while for the high permeability contrast case permeability values range between $10^{-9}$ and 
$10^{-15} \ m^2$ and $log(k_i/{k_{i+1}}) = 2$. Boundary conditions for the pressure are set as zero 
gradient along the lateral boundaries and a one dimensional pressure gradient of 50 Pa/m aligned 
with the longitudinal direction. Concentration boundary conditions are fixed single concentration 
on the inlet boundary and zero gradient on all the other sides. This type of injection is 
described by a step function.

\begin{table}[!htbp]
\centering
\begin{tabular}{ll}
\hline
\multicolumn{2}{c}{\textbf{Geostatistical parameters}}                                              
\\ \hline
Correlation function & Exponential
\\
Number of facies & 4
\\
Volumetric proportion & 25\%
\\ \hline
\multicolumn{2}{c}{\textbf{Flow parameters}}                                                        
\\ \hline
Pressure gradient [Pa/m] & 50 
\\ \hline
\multicolumn{2}{c}{\textbf{Transport parameters}}                                                   
\\ \hline
Fixed inlet concentration [-] & 1 
\\ \hline
\end{tabular}
\caption{Parameters kept constant throughout the simulations.}
\label{tab:constParam}
\end{table}
The simulation workflow is divided into three steps
\begin{itemize}
	\item geostatistics: the permeability domain is generated using the truncated 
pluri-Gaussian algorithm;
	\item flow: equation \ref{eq:Darcy} is solved with the prescribed boundary conditions and provides 
the steady state flow field;
	\item transport: advection-dispersion transient model is solved with continuous injection 
for each simulation time step.  
\end{itemize}

The simulations are run within the open-source \of-based library SECUReFoam \cite{SECUReFoam} 
which includes the \texttt{setRandomField} utility for truncated pluri-Gaussian simulations, 
\texttt{simpleDarcyFoam} and \texttt{adaptiveScalarTransportFoam} solvers for flow and transport 
simulations.
Most of the simulations were run in parallel on 96 cores split between 8 HPC nodes. An adaptive 
time step tied to the Courant number was implemented together with an automatic check on the 
section-average outlet concentration value which stopped the transport simulation when a value of 
0.99 on the outlet boundary was reached. In this setting, the overall simulation time ranges 
between 1 and 7 hours depending on the permeability contrast adopted, with high contrast cases 
being characterised by larger CPU costs. Transport simulation are the most expensive of the three 
simulation steps, accounting for between the 70 and 95 \% of the computational time, depending on 
the low or high permeability contrast setting.

\section{Results}
The results presented in this section aim to assess the impact of the parameters related to the 
PGS fields on solute transport processes. To this end, first, we first compare the PDFs of 
velocity point values obtained in the considered fields. Then, we move to the analysis of the 
transport simulations and we provide 
a qualitative assessment of the variability exhibited by results obtained from realisations of the 
conductivity fields generated with identical geostatistical parameters (section \ref{sec:realvar}). 
Finally, we analyse the impact of three physical parameters on PDFs, namely permeability contrast 
(section \ref{sec:PermCont}), the longitudinal correlation length used to generate the fields 
(section \ref{sec:CL}) and Péclet number (section \ref{sec:Peclet}). 
By testing transport behaviour in settings with increasing longitudinal correlation lengths and 
permeability contrast between facies, we aim to:
\begin{itemize}
	\item assess the sensitivity of transport behaviour and the onset of non-Fickian transport 
	with respect to the correlation length and permeability constast of the generated permeability fields;
	\item quantify the discrepancy from Fickian behaviour as a function of both correlation 
lengths and permeability contrast, considered separately.
\end{itemize}

\begin{figure}[!htbp]
\centering
  \includegraphics[width=0.45\linewidth]{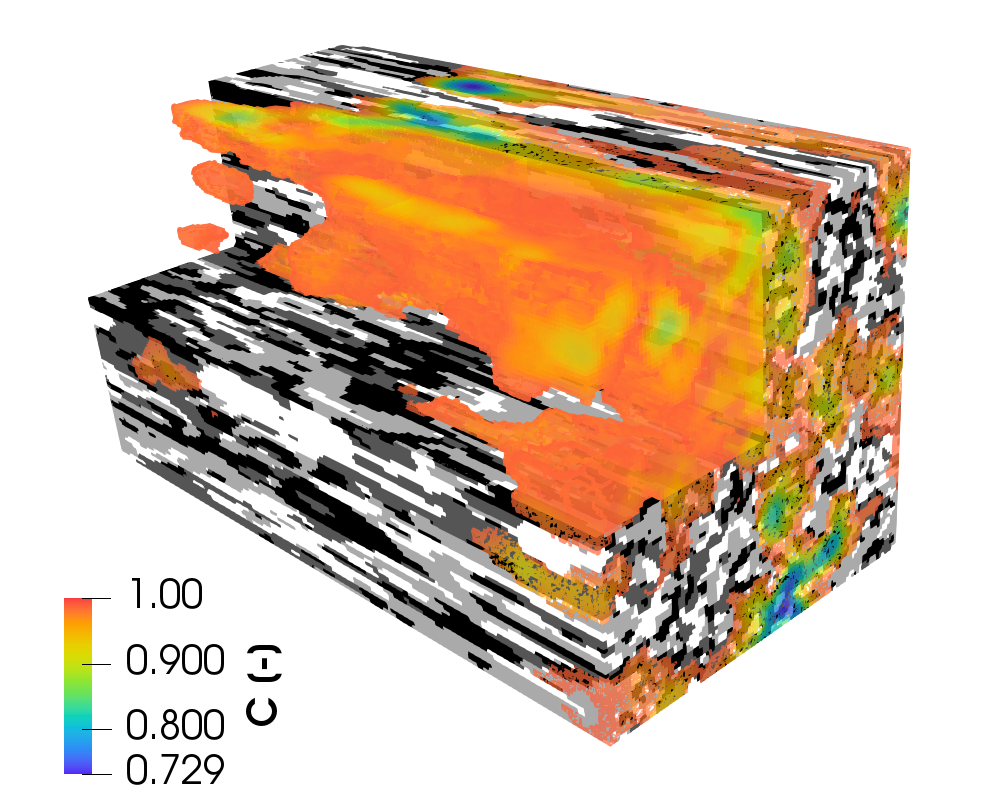}
  \includegraphics[width=0.45\linewidth]{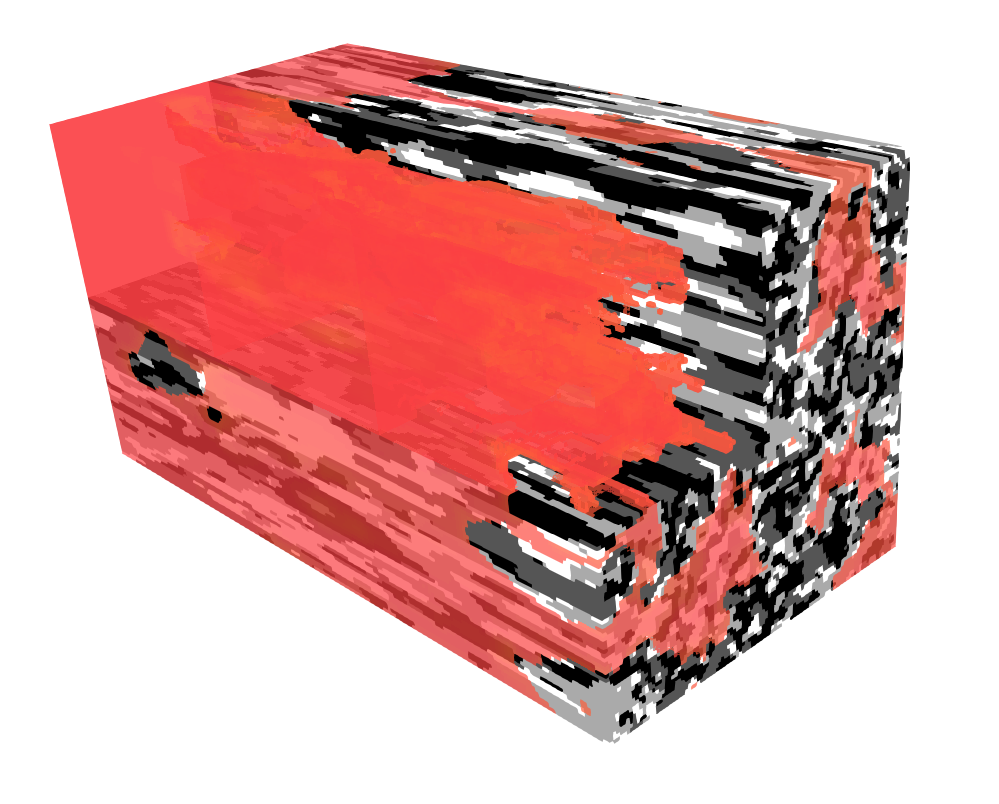}
  \label{fig:highC}
\caption{Solute plume distribution in high contrast permeability domain at late time. Domain sizes 
are 2x1x1 and correlation lengths along the three directions are set to (0.8, 0.1, 0.1). On the 
left panel it is possible to observe how low solute concentration values (0 blue - 0.99 red) are 
confined to low permeability regions ($10^{-12} \ [m^{2}]$ dark grey - $10^{-13} \ [m^{2}]$ black) 
while on the right panel high concentration values (0.99 blue - 1.00 red) are highlighted and their 
spatial distribution clearly show that saturated zones are concentrated in highly permeable regions 
($10^{-10} \ [m^{2}]$ dark grey - $10^{-11} \ [m^{2}]$ black).}
\label{fig:LowHighC}
\end{figure}

Before moving to these detailed analyses, figure \ref{fig:LowHighC} illustrates the simulation of the solute plume
at late times through a PGS field with high 
contrast permeability and characterised by longitudinal correlation length of 0.8 m. Figure 
\ref{fig:LowHighC} on the left highlights the regions where concentration values falls beneath the 
0.99 threshold while the right panel visualises the regions where concentration values fall between 
0.99 and 1. It is possible to observe that the transport of the solute is facilitated in the high 
permeable regions of the domain (white and light grey) while low permeability ones (dark grey and 
black) form a flow barrier that impede advective solute transport. 

\subsection{Velocity PDFs}
\begin{figure}[!htbp]
	\centering
	\includegraphics[width=0.45\linewidth]{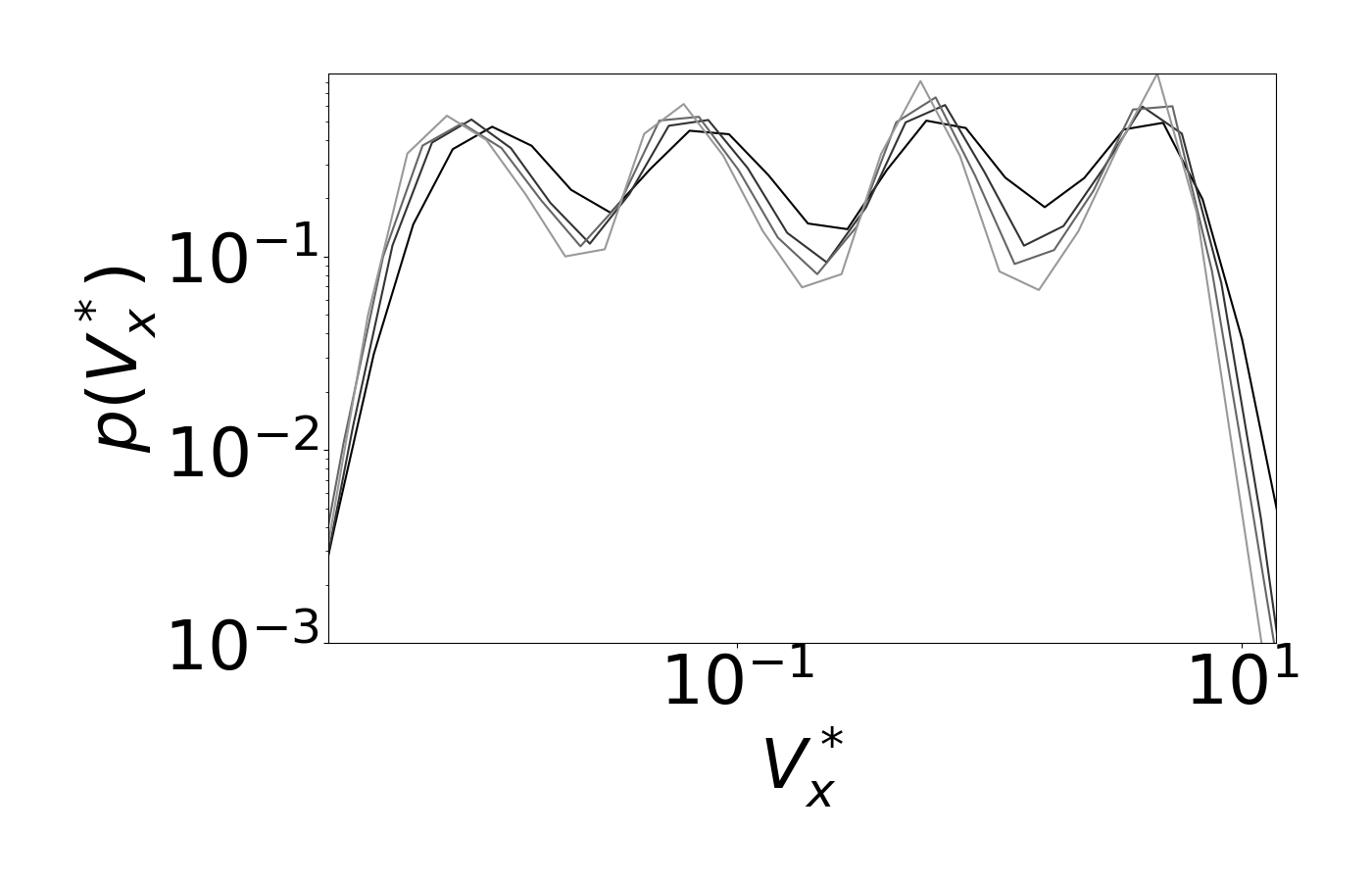}
	\includegraphics[width=0.45\linewidth]{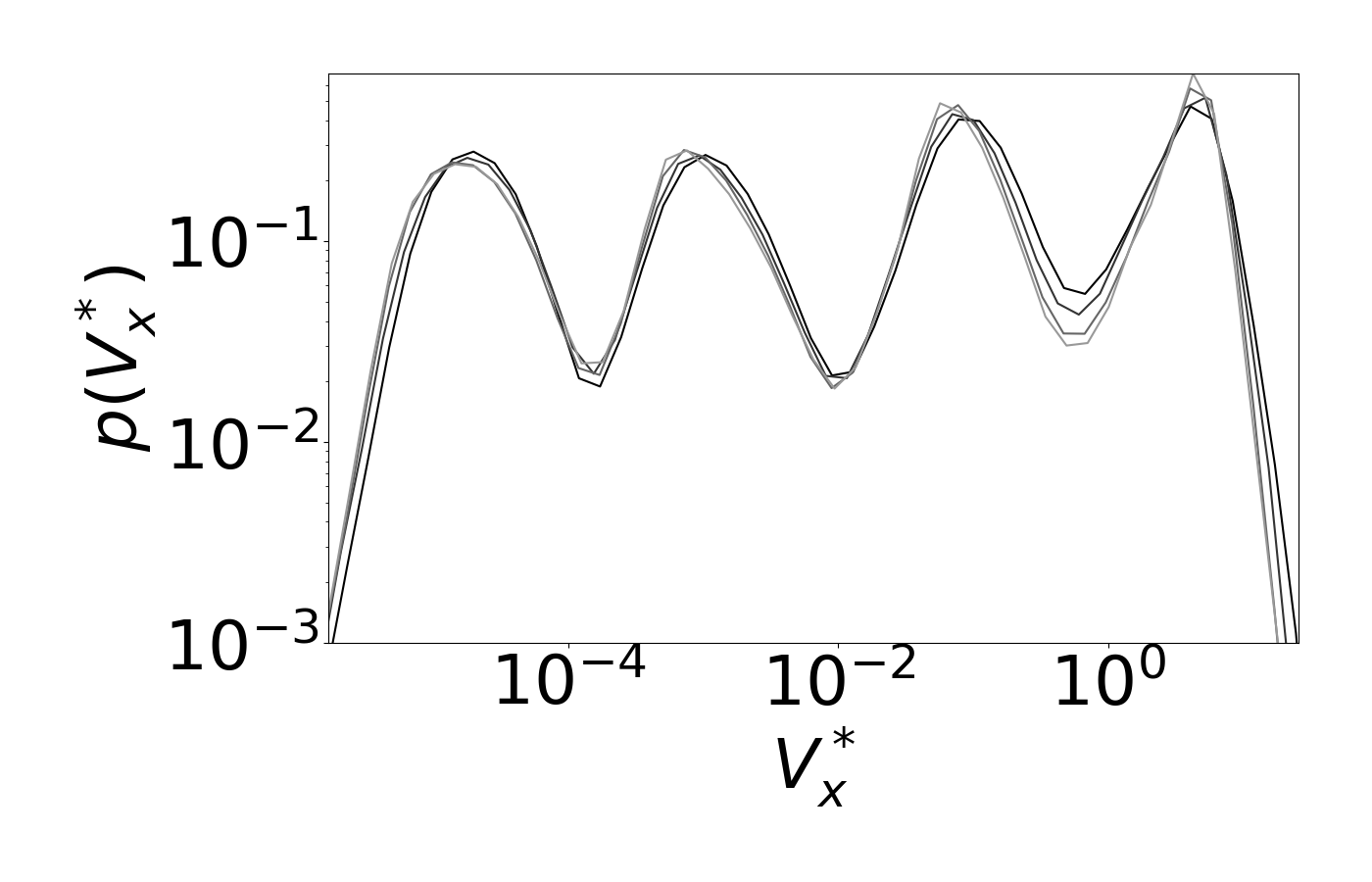}
\caption{PDFs of the longitudinal velocity component for low and high contrast. Velocity PDFs are 
shown for low (left) and high (right) permeability contrast. The correlation lengths $\lambda_x$ 
span between 0.4 (darker lines) and 1.0 (lighter lines).}
\label{fig:velPdf}
\end{figure}

\begin{figure}[!htbp]
	\centering
	\includegraphics[width=0.45\linewidth]{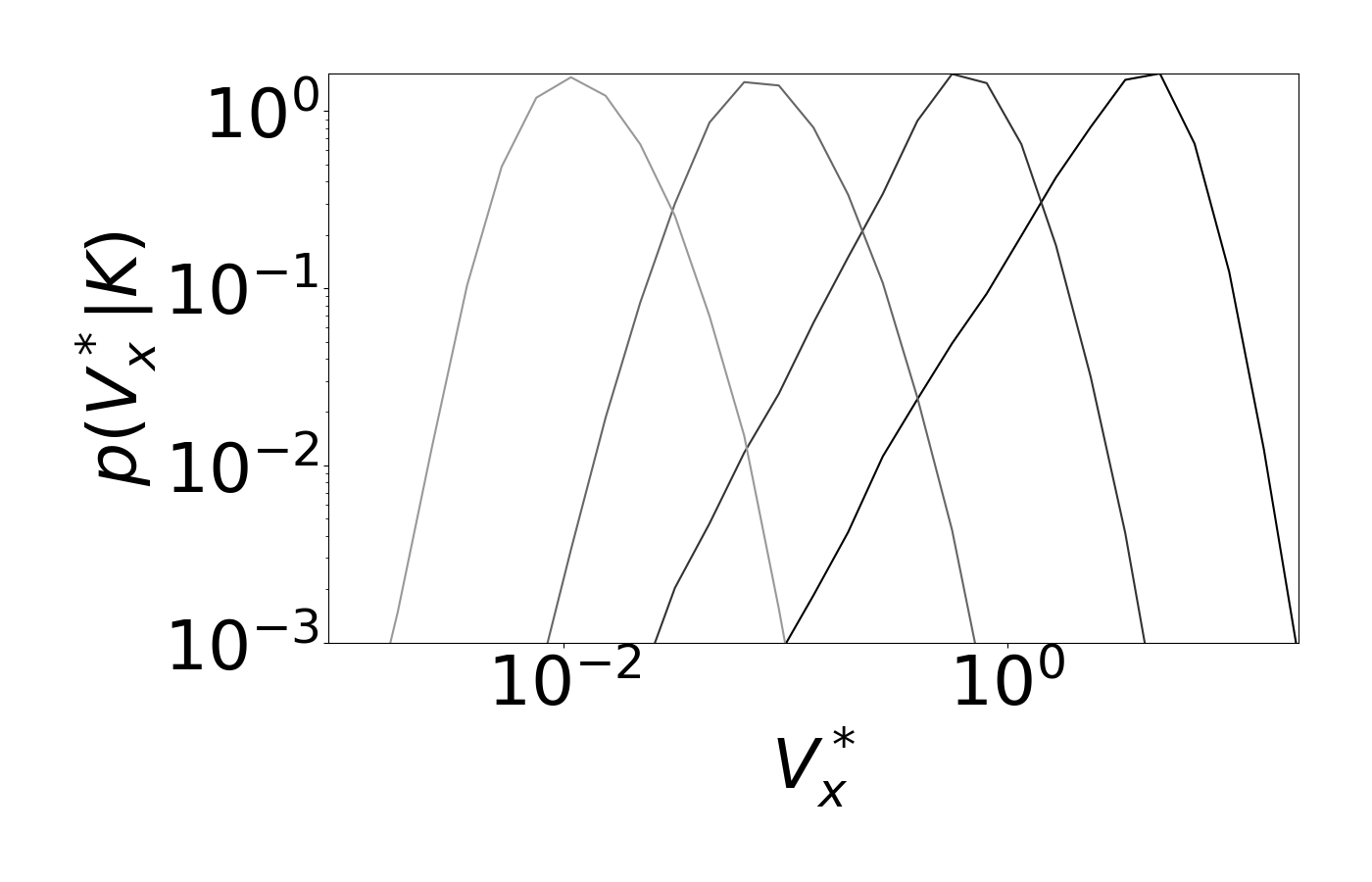}
	\includegraphics[width=0.45\linewidth]{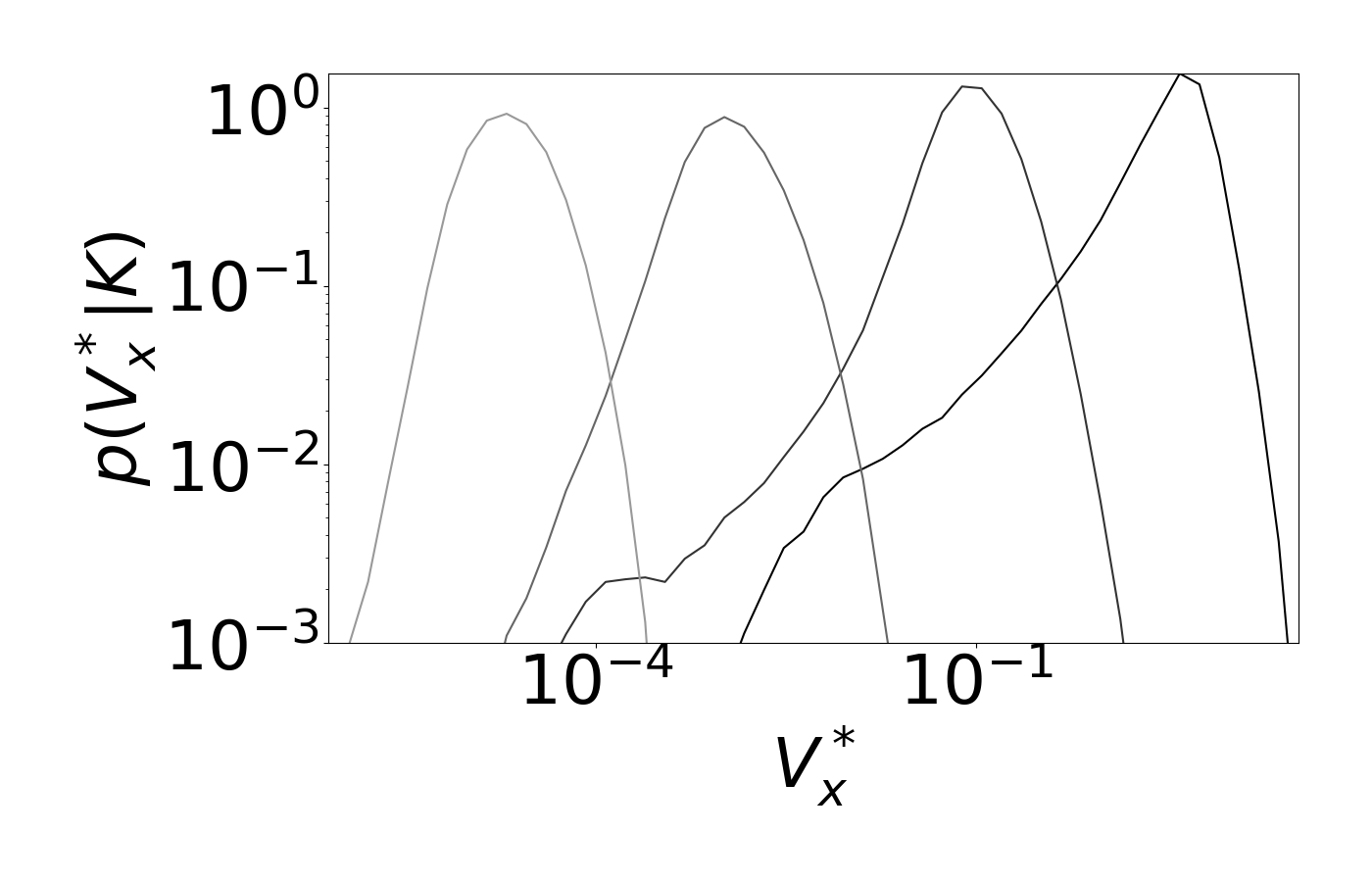}
    \caption{Conditional PDFs of the longitudinal velocity values in low (left) and high (right) 
permeability contrast. Results are shown for correlation lengths $\lambda_x = 0.4$. The curves are shown with different colors depending on the facies permeability, i.e., lighter colors correspond to low permeability and darker colors to high permeability media.}
\label{fig:velPdfCond}
\end{figure}

The velocity PDF has a direct influence on non-Fickian transport features 
\cite{comolli2019}. The PDFs of point velocity values with increasing 
longitudinal correlation lengths $\lambda_x$ are shown in figure \ref{fig:velPdf}. The longitudinal velocity $V_x$ normalised 
with the average longitudinal velocity is reported on the horizontal axis of figures 
\ref{fig:velPdf} and \ref{fig:velPdfCond} as $V^*_x$. The vertical axis of figure \ref{fig:velPdf} 
reports the probability density distribution as a function of the longitudinal velocity 
$p({V^*_x})$. As expected, the velocity distributions in figure \ref{fig:velPdf} are comparable for 
different correlation lengths as they all show four peaks of similar height corresponding to the 
four facies that populate the domain. However, as the correlation length increases, the peaks 
become sharper reflecting the formation of preferential flow paths where velocities are lumped 
around the mean velocity of a given facies, each corresponding to a mode of the distributions 
(figure \ref{fig:velPdfCond}). This result also indicates that with a decrease of $\lambda_x$ the 
distribution of velocity values progressively converges towards a uniform distribution across the 
whole range. Comparing the amplitude of the peaks in the two panels of figure \ref{fig:velPdf} we 
observe that as the permeability contrast increases the four modes of the distribution appear more 
distinct for high contrast than for low contrast. Note also that the high contrast distribution 
spans a much wider interval of velocity values as compared to the low contrast one.

Figure \ref{fig:velPdfCond} reports the conditional PDFs $p(V^*_x \vert k=k_i)$, where each 
distribution considers only longitudinal velocities values computed in cells associated with a 
given facies ($i = 1 \ldots 4$). Velocities in highly permeable regions show an asymmetric 
distribution, characterised by a pronounced peak and a leftward tail. Conversely, velocity values 
observed in the low permeability regions tend to assume a symmetric and compact distribution. This 
distinct behavior is particularly evident in high contrast media. This means that high-permeability 
regions may feature a broad distribution of velocity values because of the overall connectivity of 
the field. Highly connected regions give rise to fast channels in formations featuring large values 
of $k$ but poorly connected regions may also involve high-permeability cells. 

\subsection{Variability of transport behaviour across multiple realizations} \label{sec:realvar}
\begin{figure}[!htbp]
	\centering
	\includegraphics[width=0.45\linewidth]{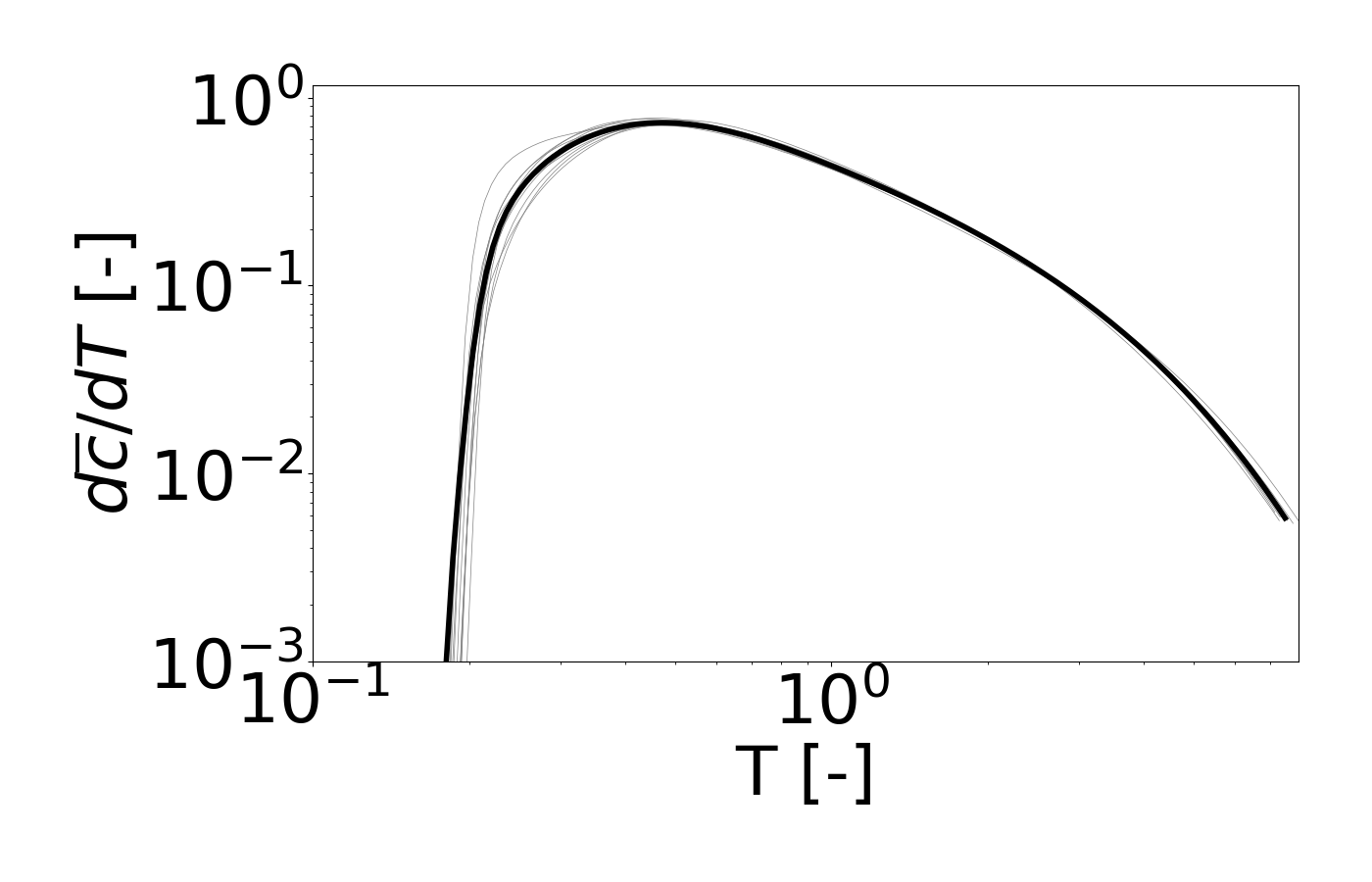}
	\includegraphics[width=0.45\linewidth]{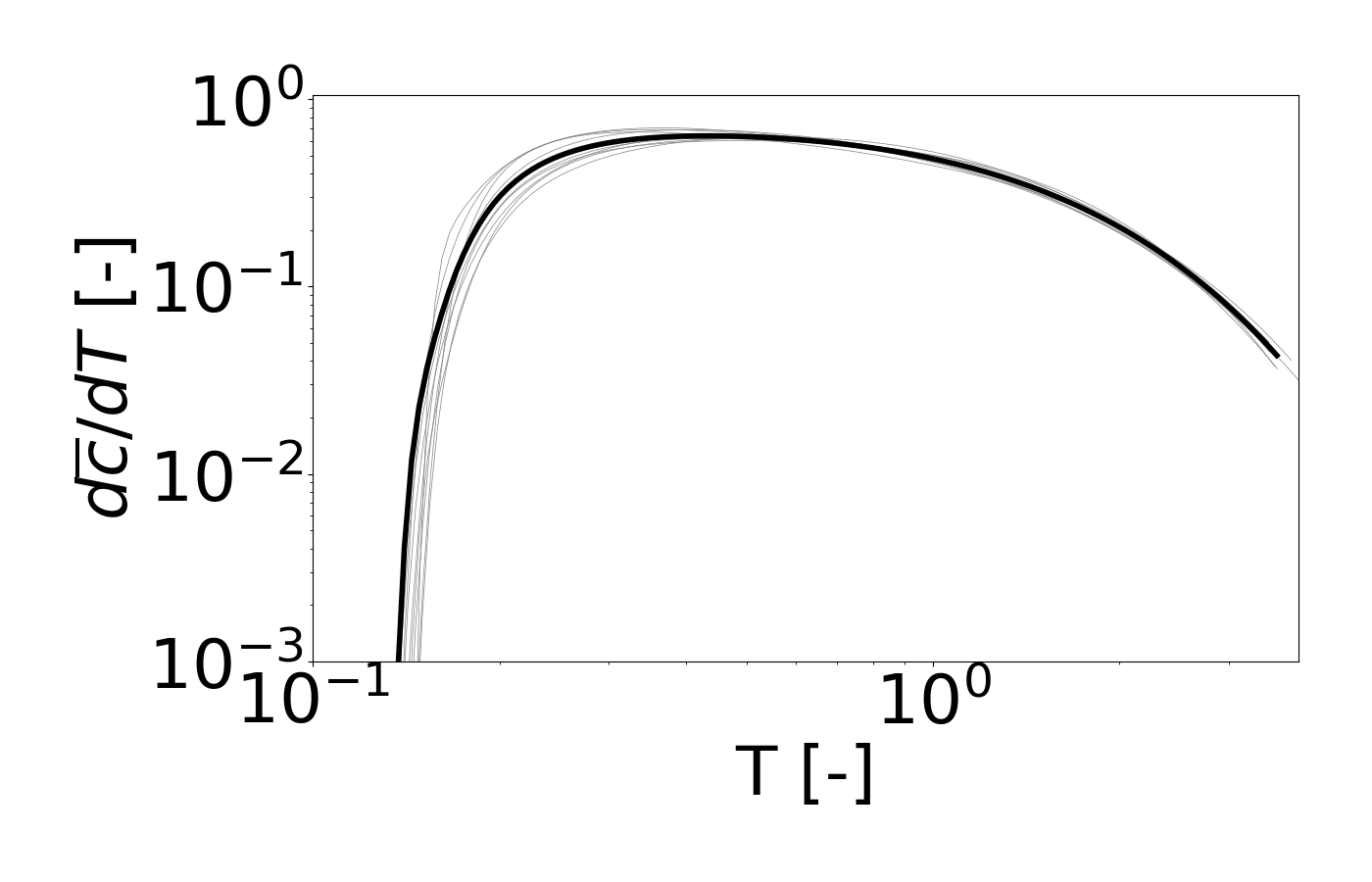}
	\caption{PDFs of solute arrival times associated with 10 realisations with the same 
correlation length ($\lambda_x = 0.8 m$). Low and high permeability contrast on the left and right 
side of the panel respectively.}
	\label{fig:realisVar}
\end{figure}

Figure \ref{fig:realisVar} displays the overlap of the PDF of the solute arrival times, obtained 
taking the time derivative of the BTC ($d\bar{c}/dT$) obtained from 10 realisations of permeability 
fields generated with the same geostatistical parameters. The observed variability tends to be 
greater for early times while at later times the different realizations attain similar values. 
This behaviour is the result of the solute exploring more broadly the facies' heterogenities as the 
solute fill the whole domain. Although the outlined behaviour does not show qualitatively relevant 
differences between low (left panel) and high (right panel) permeability contrast, for high 
permeability contrast the spread between the first part of the curves is greater than the spread 
for low permeability contrast. Because in the following we focus on the assessment of the 
macroscopic response of the system and the departure from a Fickian macrodispersive model, we deem 
a single realisation to be representative of the response of the system to various combinations of 
the investigated parameters.  

\subsection{Effect of permeability contrast} \label{sec:PermCont}
\begin{figure}[!htbp]
	\centering
	\includegraphics[width=0.45\linewidth]{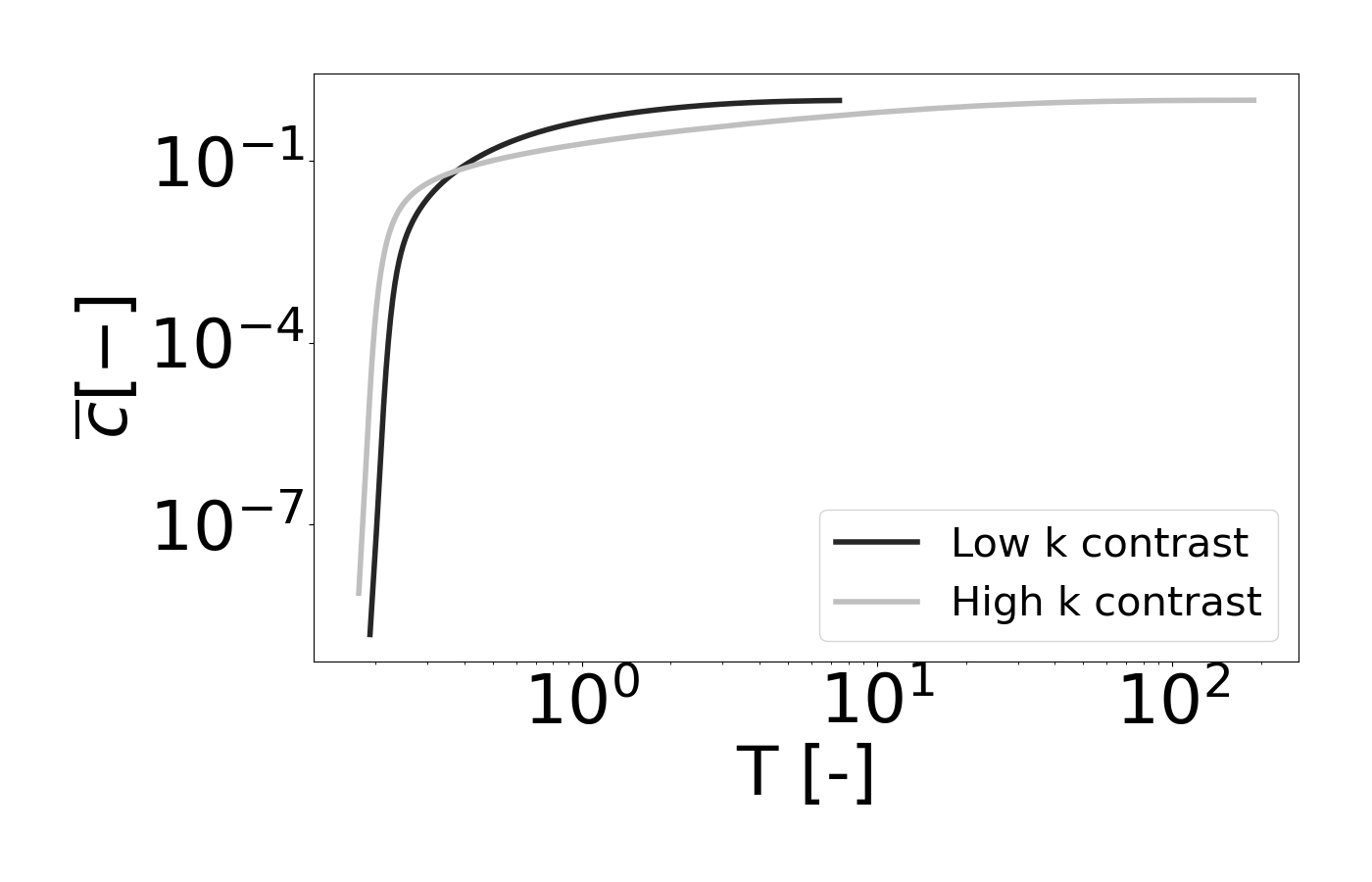}
	\includegraphics[width=0.45\linewidth]{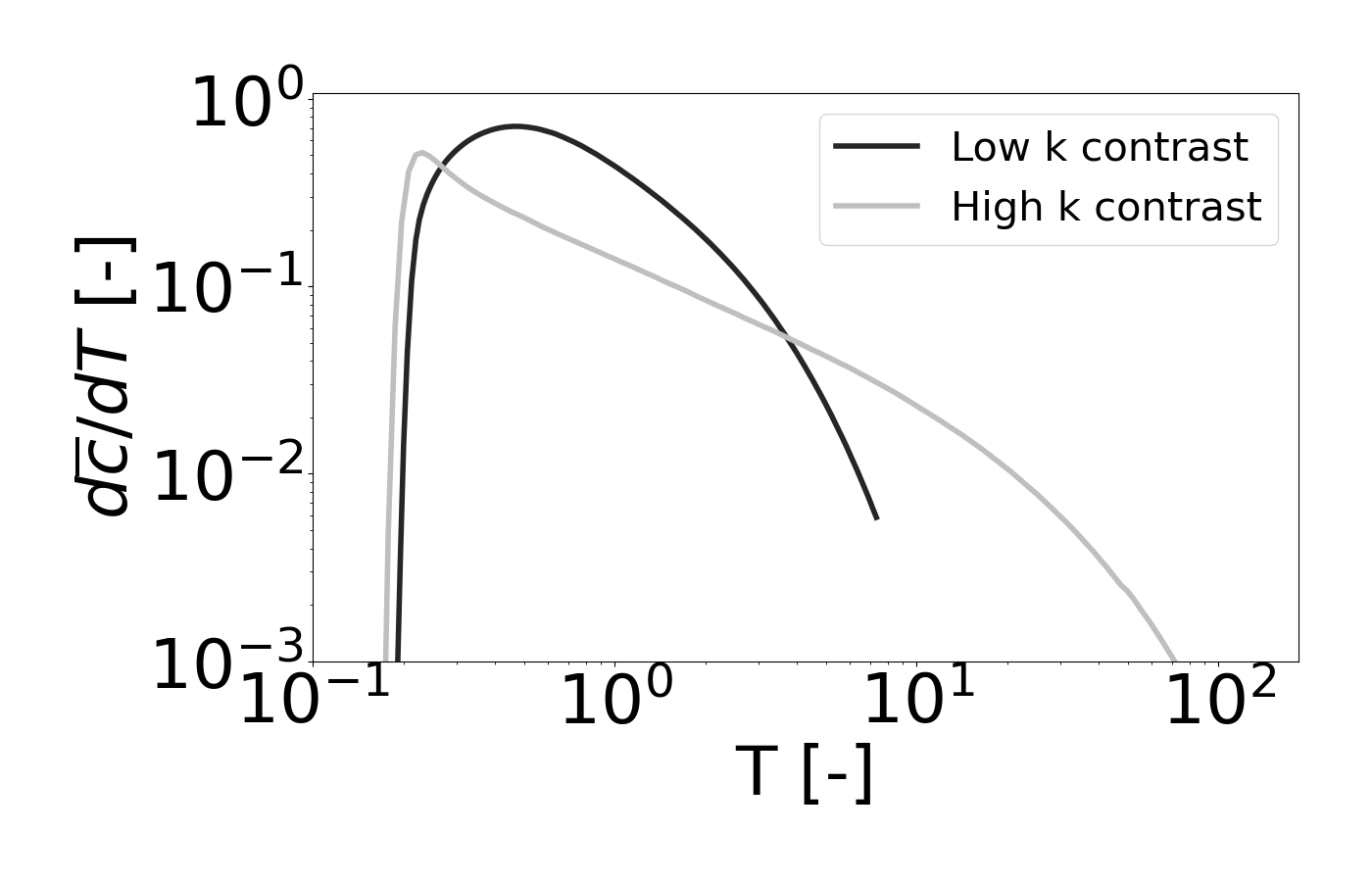}
	\includegraphics[width=0.45\linewidth]{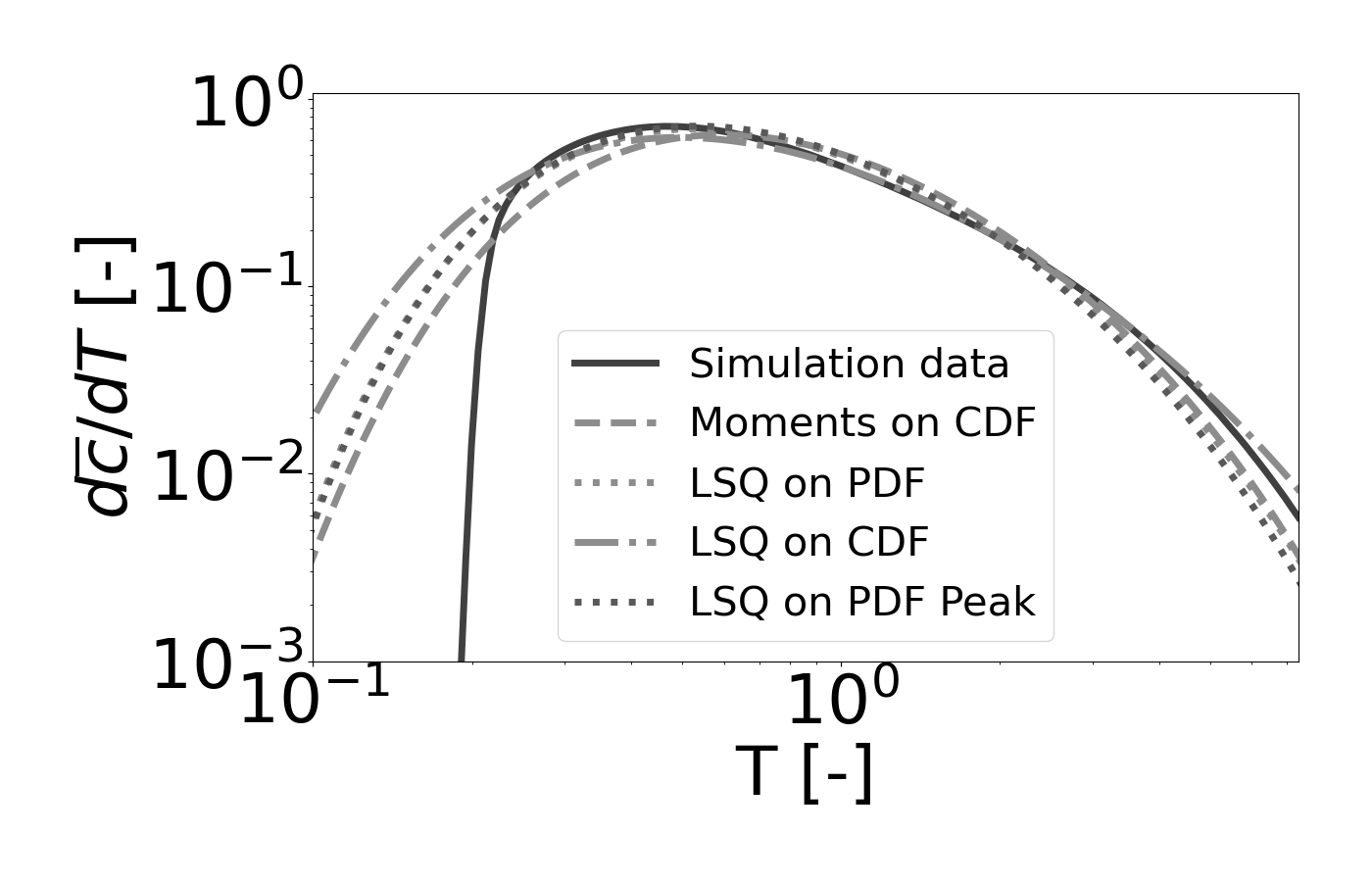}
	\includegraphics[width=0.45\linewidth]{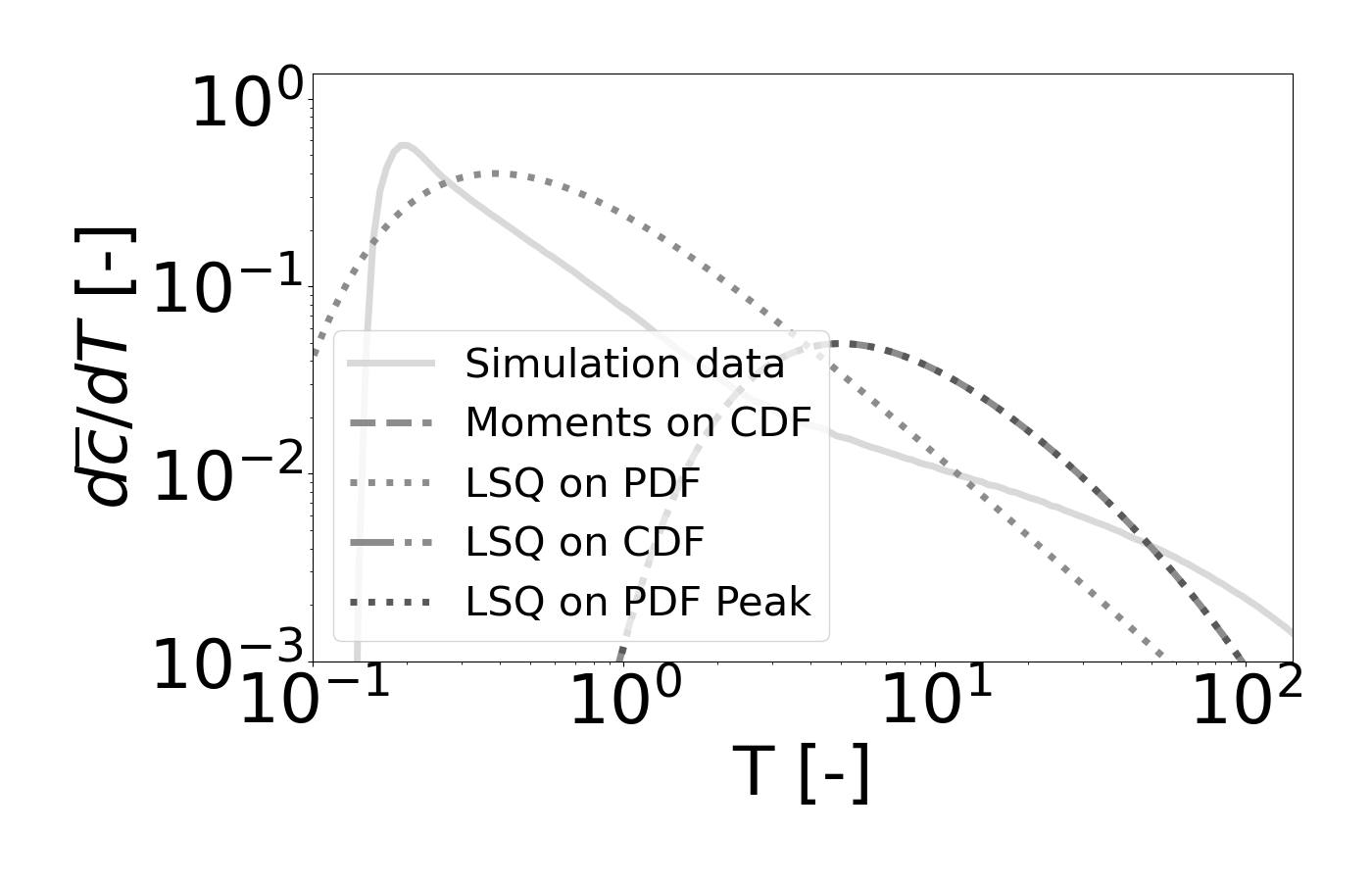}
\caption{Top panels: breakthrough curves (left) and their time derivatives (right) simulated on identical geological 
structure with low and high permeability contrast between the facies. Bottom panels: the curves are overlapped 
with the corresponding Inverse Gaussian approximations via least square (LSQ) or moments' method 
estimation in low (left) and high (right) permeability contrast case.}
\label{fig:LowHighContrast}
\end{figure}

Figure \ref{fig:LowHighContrast} illustrates the effect of the permeability contrast between facies 
on transport, by comparing the results of transport simulations performed on two geological domains 
with identical arrangement but assuming low and high permeability contrast. Geostatistical, flow and 
transport parameters relative to the results in figure 
\ref{fig:LowHighContrast} are shown in table \ref{tab:paramValues}. 

The simulated BTCs (i.e., CDFs of the solute arrival times) are shown on top left of figure 
\ref{fig:LowHighContrast} while on top right of figure \ref{fig:LowHighContrast} the corresponding 
time derivative are shown, these latter corresponding to the PDFs of arrival times. A Fickian model 
based on the inverse Gaussian distribution yields a reasonable fitting of the numerical data for 
the low contrast simulation, where the difference between one facies' permeability and another 
remains within one order of magnitude (see figure \ref{fig:LowHighContrast}, bottom left). In this 
case the match between the numerical simulation and the Inverse Gaussian distribution is 
satisfactory especially for the peak and the right tail of the distribution. Conversely, early 
arrival times are not well approximated by the Fickian model. In the low contrast case the results 
obtained with different estimation methods are self-consistent, i.e. least squares and moments 
methods yield similar outcomes. If the contrast in permeability increases, the Inverse Gaussian 
distribution cannot match the simulated data (figure \ref{fig:LowHighContrast}, bottom right), 
regardless of the method used to estimate its parameters (Least Square or Moments method). 

In summary, figure \ref{fig:LowHighContrast} suggests that as the permeability contrast increases, 
the evolution of the solute concentration shows significant departure from the Fickian model.

\begin{table}[!htbp]
\centering
\begin{tabular}{lll}
\hline
  & Low contrast & High contrast \\ \hline
\multicolumn{3}{c}{\textbf{Geostatistical parameters}} \\ \hline
$\lambda_x$ [m] & $0.8$ & $0.8$ \\
$log(k_{i+1}/{k_i}) \ [-]$ & $1$ & $2$ \\
$Var(log(k)) \ [log \ m^4]$ & $6.5$ & $26.1$ \\ \hline
\multicolumn{3}{c}{\textbf{Flow parameters}} \\ \hline
$k_{eff,x} \ [m^2]$ & $2.11 \cdot 10^{-11}$ & $1.49 \cdot 10^{-10}$ \\
$\overline{V_x} \ [m/s]$ & $1.07 \cdot 10^{-6}$ & $9.64 \cdot 10^{-6}$ \\
$Pe_x [-]$ & $21113$ & $59633$ \\ \hline
\end{tabular}
\caption{Geostatistical and flow parameters for the low and high permeability contrast fields used 
to generate results reported in Figure \ref{fig:LowHighContrast}. The flow parameters were defined 
by equations \eqref{eq:Pe} and \eqref{eq:kEff}.}
\label{tab:paramValues}
\end{table}


A detailed analysis of these results is shown in Table \ref{tab:VelDispErr}. Numerical results 
confirm that the relative error between Inverse Gaussian approximations and the numerical solution 
is significantly lower for low permeability contrast than for high permeability contrast. It is 
then possible to conclude that the level of accuracy of macrodispersion models in capturing 
transport behaviour decreases with the permeability contrast. Table \ref{tab:VelDispErr} reports 
the values of macrodispersion parameters computed through approximation \eqref{eq:ExpDmac} 
(considered as a reference value) and compare them with the the estimated ones. Transport 
parameters estimations are closer to the reference values for the low contrast if compared to the 
high contrast cases. Moreover the estimates obtained through least squares in the high contrast 
case are generally affected by large confidence bounds (i.e., they are indicated in italic) 
indicating that the estimated values cannot be considered as reliable.


\begin{table}[!htbp]
\centering
\tabcolsep=0.09cm
\begin{tabular}{cllllll}
\multicolumn{1}{l}{} & \multicolumn{3}{c}{\textbf{Low contrast}} & \multicolumn{3}{c}{\textbf{High 
contrast}} \\
\textit{} & \multicolumn{1}{c}{$\bar{V}_x [m/s]$} & \multicolumn{1}{c}{$D_{mac} [m^2/s]$} & 
\multicolumn{1}{c}{$e [\%]$} & \multicolumn{1}{c}{$\bar{V}_x [m/s]$} & \multicolumn{1}{c}{$D_{mac} 
[m^2/s]$} & \multicolumn{1}{c}{$e [\%]$} \\ \hline
\textit{Reference solution} & $1.07 \cdot 10^{-6}$ & $8.57 \cdot 10^{-7}$ & $-$ & $9.64 \cdot 
10^{-6}$ & $7.71 \cdot 10^{-6}$ & $-$ \\
\textit{Method 1} & $6.94 \cdot 10^{-7}$ & $4.97 \cdot 10^{-7}$ & $16.20$ & $7.03 \cdot 10^{-7}$ & 
$1.59 \cdot 10^{-6}$ & $79.5$ \\
\textit{Method 2} & $2.08 \cdot 10^{-7}$ & $1.81 \cdot 10^{-7}$ & $10.71$ & $\mathit{3.79 \cdot 
10^{-5}}$ & $\mathit{8.79 \cdot 10^{-7}}$ & $\mathit{48.1}$ \\
\textit{Method 3} & $6.50 \cdot 10^{-7}$ & $6.09 \cdot 10^{-7}$ & $11.94$ & $6.36 \cdot 10^{-7}$ & 
$2.33 \cdot 10^{-6}$ & $126.7$ \\
\textit{Method 4} & $2.42 \cdot 10^{-7}$ & $1.13 \cdot 10^{-7}$ & $11.93$ & $\mathit{8.95 \cdot 
10^{-8}}$ & $\mathit{3.74 \cdot 10^{-5}}$ & $\mathit{49.2}$ \\ \hline
\end{tabular}
\caption{Simulated ($Reference \ solution$) and estimated ($Methods \ 1-4$) values for average 
Darcy velocity $\bar{V}_x$, macrodispersion $D_{mac}$ and relative breakthrough error $e$ values. 
The $reference \ solution$ values represent the average longitudinal velocity and the nominal 
macrodispersion as from equation \eqref{eq:ExpDmac}. $Method \ 1$ is the moments method, $Method \ 
2, \ 3$ and $4$ correspond to least squares method applied to the simulated PDF, CDF and the PDF 
peak data. As a result of the unsuitability of the Inverse Gaussian model to the describe the PDF 
in high permeability contrast scenarios, some values (\textit{italic}) are characterised by 
extremely large standard deviations ($\sigma>10^6$).}
\label{tab:VelDispErr}
\end{table}

\subsection{Correlation length} \label{sec:CL}

\begin{figure}[!htbp]
	\centering
	\includegraphics[width=0.45\linewidth]{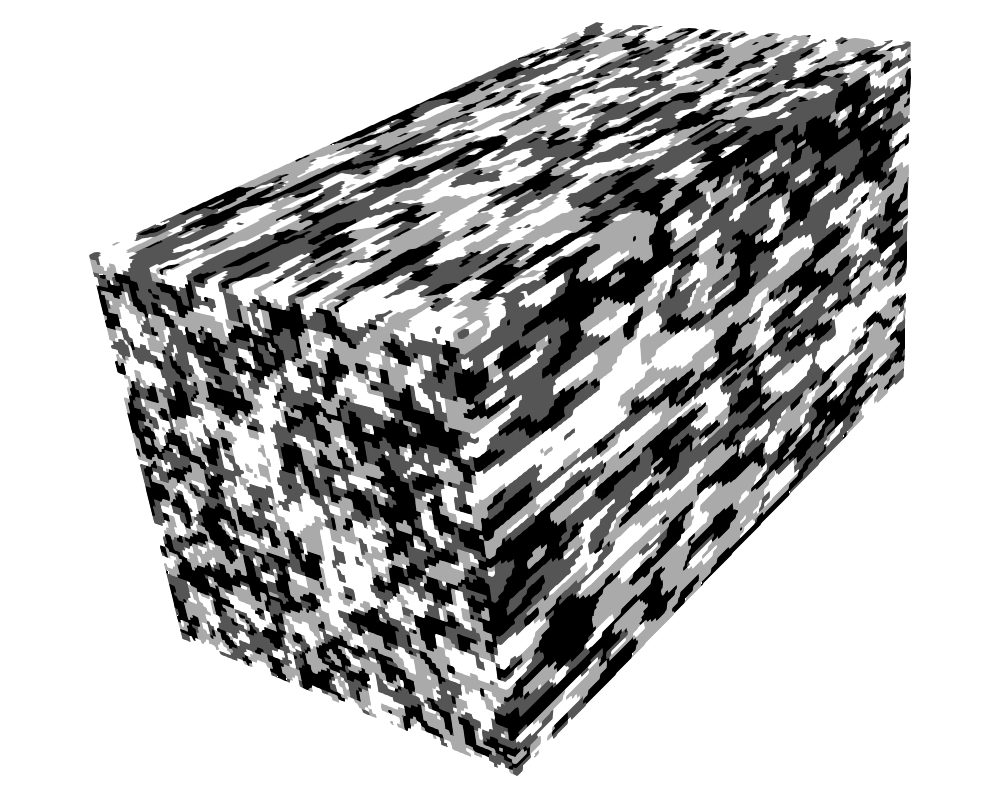}
	\includegraphics[width=0.45\linewidth]{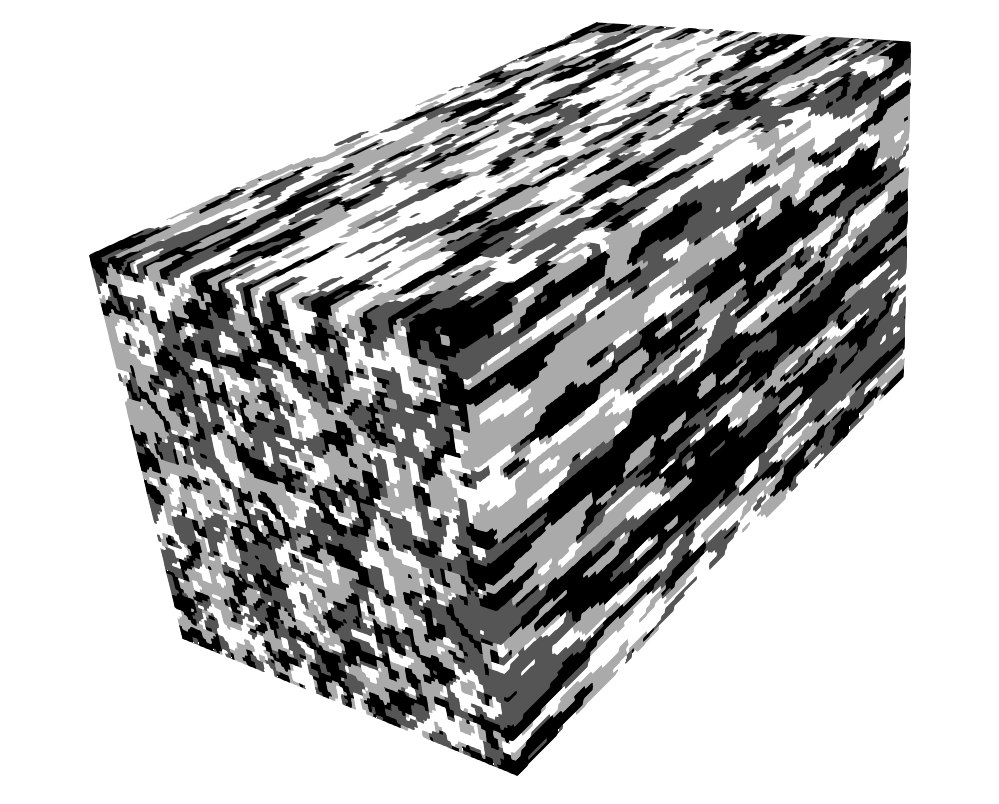}
	\includegraphics[width=0.45\linewidth]{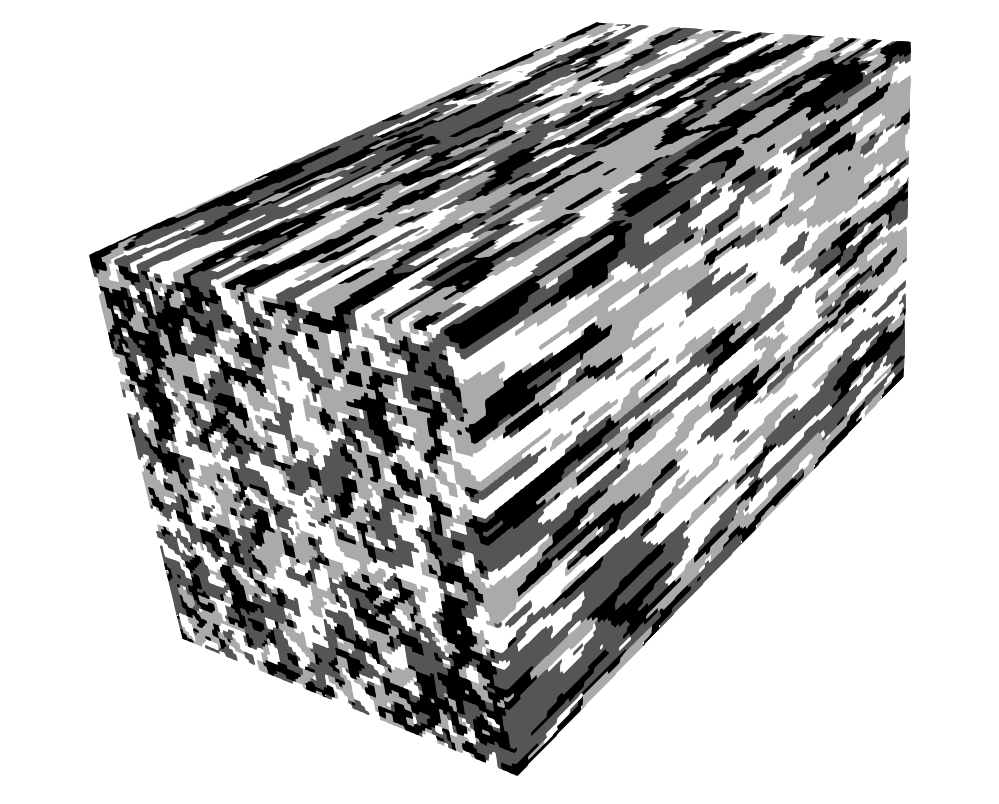}
	\includegraphics[width=0.45\linewidth]{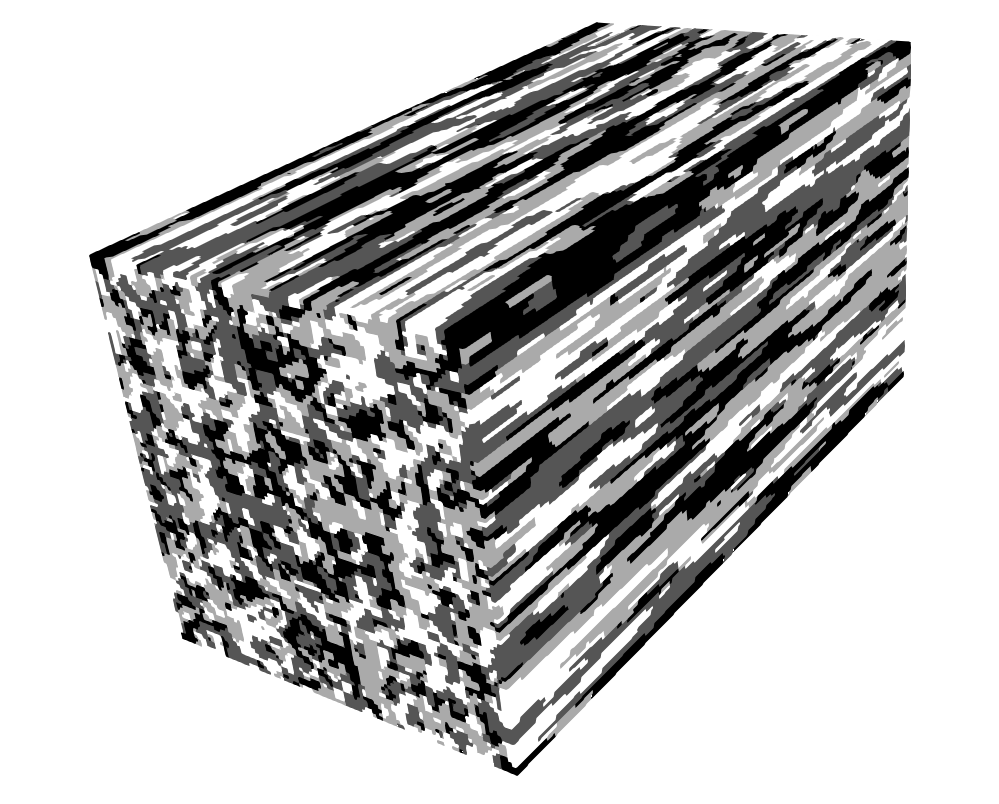}
\caption{Truncated pluri-Gaussian permeability fields with increasing longitudinal correlation 
length $\lambda_x$. Clockwise order from top left panel: $\lambda_x = 0.4$, $\lambda_x = 0.6$, 
$\lambda_x = 0.8$, $\lambda_x = 1.0$.}
\label{fig:CorrLx}
\end{figure}

\begin{figure}[!htbp]
  \includegraphics[width=0.45\linewidth]{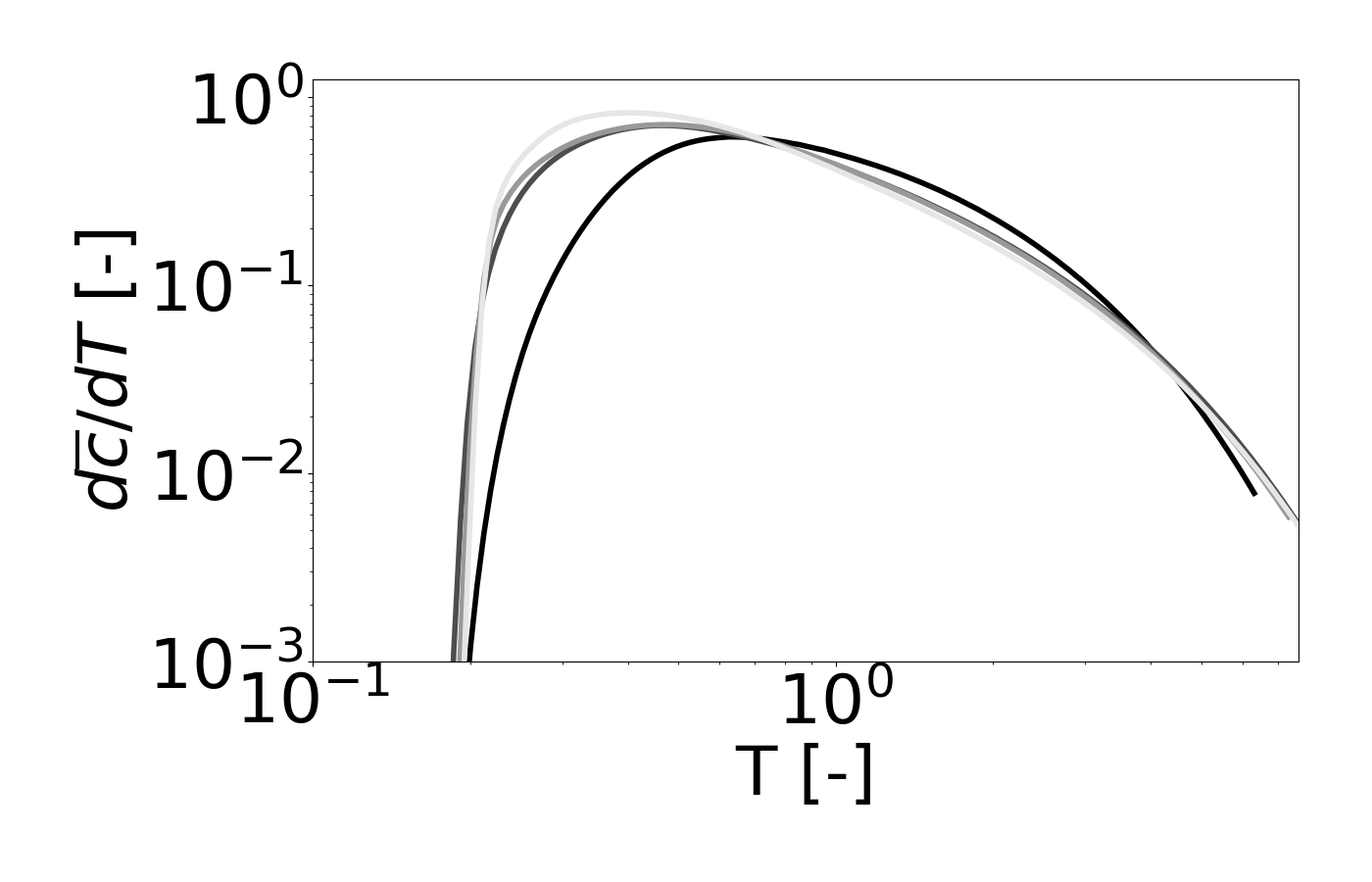}
  \includegraphics[width=0.45\linewidth]{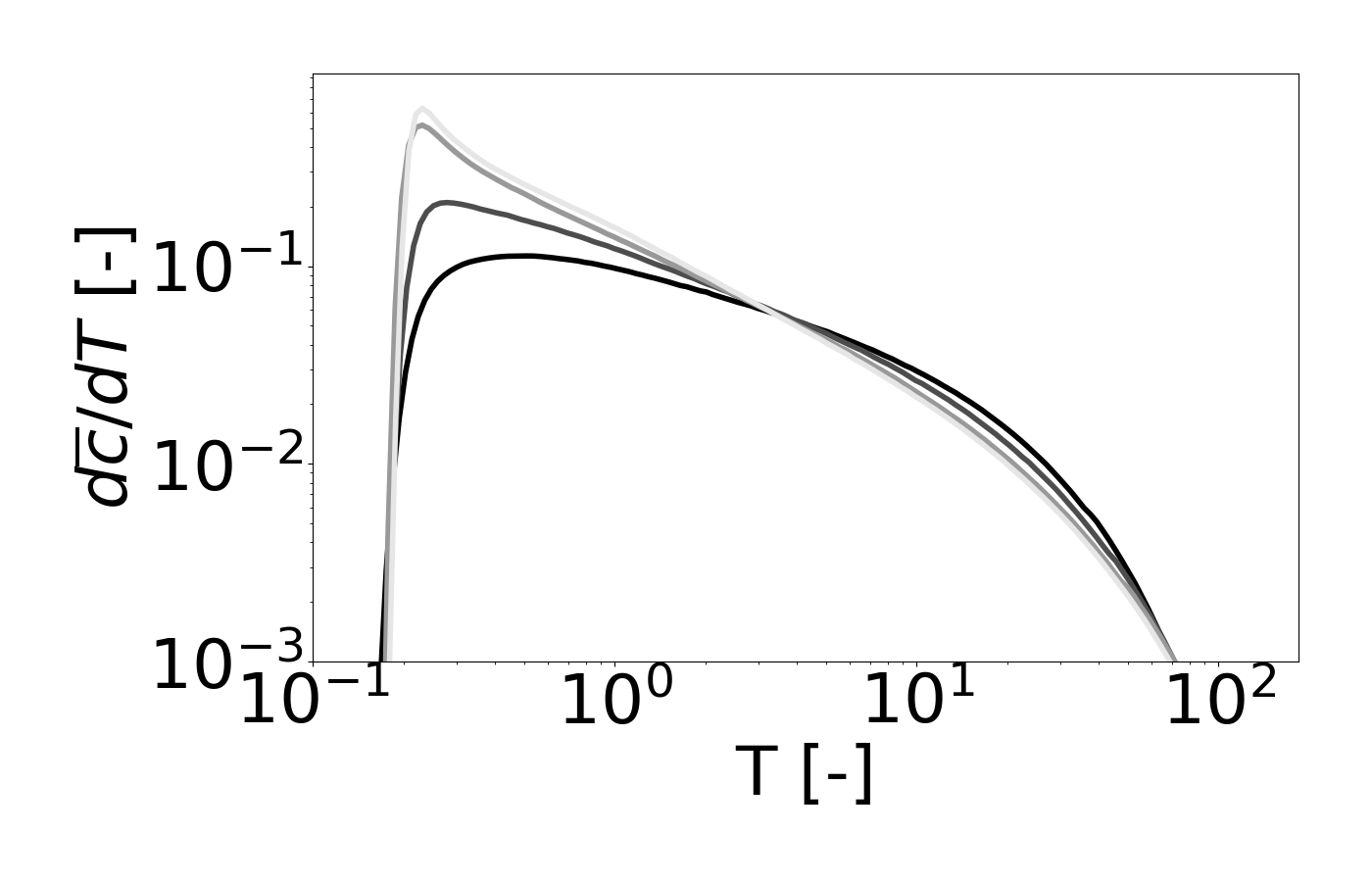}
\caption{Arrival time PDFs computed as $d\bar{c}/dT$ for low (left panel) and high (right panel) permeability contrast 
for as a function of the longitudinal correlation length $\lambda_x$. From the darkest to the 
lightest curve the longitudinal correlation length $\lambda_x=1$ increases with evenly spaced 
interval from 0.4 to 1.}
\label{fig:LowHighKbtc}
\end{figure}

Transport simulations are performed on PGS domains sharing comparable geostatistical parameters (table 
\ref{tab:paramValuesLxLC}) while increasing longitudinal correlation lengths (figure 
\ref{fig:CorrLx}). These provide interesting insights into the transition from Fickian to anomalous 
transport in relation to the connectivity degree of the sediment structure (figure 
\ref{fig:LowHighKbtc}). We emphasize here that the correlation length mentioned here is the one 
employed to generate the continuous Gaussian random fields which are then employed to generate the 
conductivity fields (see Figure \ref{fig:PGSmethod}). This length can be interpreted as the characteristic length over which facies' transitions are observed.

Figure \ref{fig:LowHighKbtc} displays the PDFs of the solute arrival times obtained for a range of values 
assigned to $\lambda_x$. As the longitudinal correlation length increases, the magnitude of the 
peak value increases and the peak shifts towards earlier arrival times. This can be explained 
observing that, with increasing correlation in the longitudinal direction, the connectivity between 
highly permeable facies favours the formation of fast channels where advection prevails over 
diffusion thus leading to early arrivals. This effect is more evident for the high contrast 
scenario (right side of figure \ref{fig:LowHighKbtc}) and is reflected by the solute arrivals PDFs 
trends: as the domain connectivity increases, the initial concentration peak rises while the 
central segment of the curve highlights a power law response.

Tables \ref{tab:paramValuesLxLC} and \ref{tab:paramValuesLxHC} report relevant geostatistical and 
flow simulation parameters which are required to interpret the results of the average Darcy 
velocity and macrodispersion estimation process provided in tables \ref{tab:VelDispErrLC} and 
\ref{tab:VelDispErrHC}. As a result of the emergence of preferential flow-paths, the effective 
permeability shows a positive trend for increasing correlation lengths.

\begin{table}[!htbp]
\centering
\begin{tabular}{lllll}
\hline
& $\bs{\lambda_x = 0.4}$ & $\bs{\lambda_x = 0.6}$ & $\bs{\lambda_x = 0.8}$ & $\bs{\lambda_x = 1.0}$ 
\\ \hline
\multicolumn{5}{c}{\textit{Geostatistical parameters}} \\ \hline
$log(k_i/{k_{i+1}})$ & $1$ & $1$ & $1$ & $1$ \\
$Var(log(k)) \ [log \ m^4]$ & $6.7$ & $6.6$ & $6.6$ & $6.6$ \\ \hline
\multicolumn{5}{c}{\textit{Flow parameters}} \\ \hline
$k_{eff,x} \ [m^2]$ & $1.75 \cdot 10^{-11}$ & $1.99 \cdot 10^{-11}$ & $2.14 \cdot 10^{-11}$ & $2.27 
\cdot 10^{-11}$ \\
$P e_x [-]$ & $3508$ & $5972$ & $8571$ & $11355$ \\ \hline
\end{tabular}
\caption{Geostatistical and flow parameters for low permeability contrast simulations. The flow 
parameters are defined by equations \eqref{eq:Pe} and \eqref{eq:kEff}.}
\label{tab:paramValuesLxLC}
\end{table}

\begin{table}[!htbp]
\centering
\begin{tabular}{lllll}
& $\bs{\lambda_x = 0.4}$ & $\bs{\lambda_x = 0.6}$ & $\bs{\lambda_x = 0.8}$ & $\bs{\lambda_x = 1.0}$ 
\\ \hline
\multicolumn{5}{c}{\textit{Geostatistical parameters}} \\ \hline
$log(k_i/{k_{i+1}})$ & $2$ & $2$ & $2$ & $2$ \\
$Var(log(k)) \ [log \ m^4]$ & $26.1$ & $26.2$ & $25.9$ & $26.2$ \\ \hline
\multicolumn{5}{c}{\textit{Flow parameters}} \\ \hline
$k_{eff, x} \ [m^2]$ & $1.50 \cdot 10^{-10}$ & $1.76 \cdot 10^{-10}$ & $1.93 \cdot 10^{-10}$ & 
$2.05 \cdot 10^{-10}$ \\
$Pe_x [-]$ & $29953$ & $52919$ & $77127$ & $102719$ \\ \hline
\end{tabular}
\caption{Geostatistical and flow parameters' for high permeability contrast simulations. The flow 
parameters were defined by equations \eqref{eq:Pe} and \eqref{eq:kEff} and are computed in the 
longitudinal direction of the flow. Longitudinal correlation length increases from 0.4 to 1.0.}
\label{tab:paramValuesLxHC}
\end{table}

\begin{figure}[!htpb]
	\centering
	\includegraphics[width=0.9\linewidth]{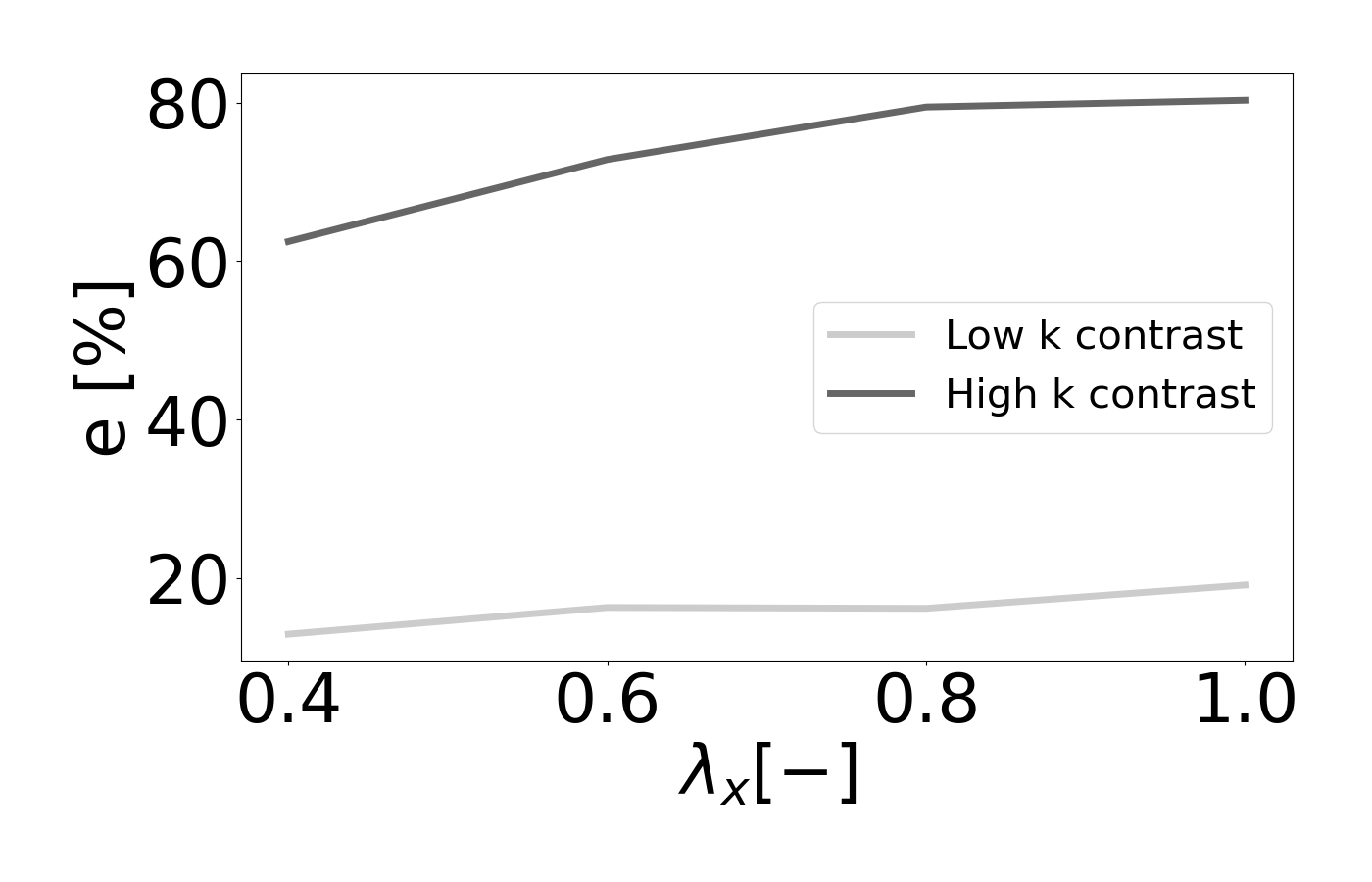}
    	\caption{Longitudinal correlation length vs relative error computed as the departure from 
Fickian model (eq. \ref{eq:RelErr}) with parameters estimated with the moments method (Method 1).}
	\label{fig:LxVsErr}
\end{figure}

Comparing the relative error computed for low and high permeability contrast cases, it is clear 
that the reliability of the Fickian model at the macroscale decreases with the permeability 
contrast (see figure \ref{fig:LxVsErr}). We also observe a mild increasing trend of the relative 
error with increasing values of the correlation length. This trend is justified by the role of the 
preferential flow-paths which facilitate fast advective flow that make overall solute behaviour anomalous.

\newgeometry{top=0.5cm, bottom=0.5cm}
\begin{sidewaystable}[!htbp]
\centering
\tabcolsep=0.11cm
\begin{tabular}{cllllllllllll}
\multicolumn{1}{l}{} & \multicolumn{12}{c}{\textbf{Low contrast}} \\
\multicolumn{1}{l}{} & \multicolumn{3}{c}{$\bs{\lambda_x = 0.4}$} & 
\multicolumn{3}{c}{$\bs{\lambda_x = 0.6}$} & \multicolumn{3}{c}{$\bs{\lambda_x = 0.8}$} & 
\multicolumn{3}{c}{$\bs{\lambda_x = 1.0}$} \\
\textit{} & \multicolumn{1}{c}{$\bar{V}_x [m/s]$} & \multicolumn{1}{c}{$D [m^2/s]$} & 
\multicolumn{1}{c}{$e [\%]$} & \multicolumn{1}{c}{$\bar{V}_x [m/s]$} & \multicolumn{1}{c}{$D 
[m^2/s]$} & \multicolumn{1}{c}{$e [\%]$} & \multicolumn{1}{c}{$\bar{V}_x [m/s]$} & 
\multicolumn{1}{c}{$D [m^2/s]$} & \multicolumn{1}{c}{$e [\%]$} & \multicolumn{1}{c}{$\bar{V}_x 
[m/s]$} & \multicolumn{1}{c}{$D [m^2/s]$} & \multicolumn{1}{c}{$e [\%]$} \\ \hline
\textit{Ref. s.} & $8.77 \cdot 10^{-7}$ & $3.51 \cdot 10^{-7}$ & $-$ & $9.95 \cdot 10^{-7}$ & $5.97 
\cdot 10^{-7}$ & $-$ & $1.07 \cdot 10^{-6}$ & $8.56 \cdot 10^{-7}$ & $-$ & $1.14 \cdot 10^{-6}$ & 
$1.14 \cdot 10^{-6}$ & $-$ \\
\textit{M. 1} & $5.39 \cdot 10^{-7}$ & $2.68 \cdot 10^{-7}$ & $12.9$ & $6.26 \cdot 10^{-7}$ & $4.54 
\cdot 10^{-7}$ & $16.3$ & $6.94 \cdot 10^{-7}$ & $4.97 \cdot 10^{-7}$ & $16.2$ & $7.60 \cdot 
10^{-7}$ & $6.15 \cdot 10^{-7}$ &  $19.1$\\
\textit{M. 2} & $1.30 \cdot 10^{-6}$ & $6.18 \cdot 10^{-8}$ & $7.3$ & $1.91 \cdot 10^{-6}$ & $1.68 
\cdot 10^{-7}$ & $11.0$ & $2.08 \cdot 10^{-6}$ & $1.81 \cdot 10^{-7}$ & $10.7$ & $2.48 \cdot 
10^{-6}$ & $2.32 \cdot 10^{-7}$ &  $16.7$\\
\textit{M. 3} & $5.12 \cdot 10^{-7}$ & $3.17 \cdot 10^{-7}$ & $28.2$ & $5.87 \cdot 10^{-7}$ & $5.55 
\cdot 10^{-7}$ & $11.8$ & $6.50 \cdot 10^{-7}$ & $6.09 \cdot 10^{-7}$ & $11.9$ & $7.10 \cdot 
10^{-7}$ & $7.60 \cdot 10^{-7}$ &  $15.8$\\
\textit{M. 4} & $2.03 \cdot 10^{-6}$ & $1.31 \cdot 10^{-7}$ & $23.7$ & $2.32 \cdot 10^{-6}$ & $1.27 
\cdot 10^{-7}$ & $13.0$ & $2.41 \cdot 10^{-6}$ & $1.13 \cdot 10^{-7}$ & $11.9$ & $2.68 \cdot 
10^{-6}$ & $1.17 \cdot 10^{-7}$ &  $16.4$\\ \hline
\end{tabular}
\caption{Simulated ($Reference \ solution$) and estimated ($Methods \ 1-4$) values for average 
Darcy velocity $\bar{V}_x$, macrodispersion $D_{mac}$ and relative breakthrough error $e$ values in 
low permeability contrast simulations. $Method \ 1$ is the moments method, $Method \ 2, \ 3$ and 
$4$ correspond to least squares method applied to the simulated PDF, CDF and the PDF peak data.}
\label{tab:VelDispErrLC}

\centering
\tabcolsep=0.11cm
\begin{tabular}{cllllllllllll}
\multicolumn{1}{l}{} & \multicolumn{12}{c}{\textbf{High contrast}} \\
\multicolumn{1}{l}{} & \multicolumn{3}{c}{$\bs{\lambda_x = 0.4}$} & 
\multicolumn{3}{c}{$\bs{\lambda_x = 0.6}$} & \multicolumn{3}{c}{$\bs{\lambda_x = 0.8}$} & 
\multicolumn{3}{c}{$\bs{\lambda_x = 1.0}$} \\
\textit{} & \multicolumn{1}{c}{$\bar{V}_x [m/s]$} & \multicolumn{1}{c}{$D [m^2/s]$} & 
\multicolumn{1}{c}{$e [\%]$} & \multicolumn{1}{c}{$\bar{V}_x [m/s]$} & \multicolumn{1}{c}{$D 
[m^2/s]$} & \multicolumn{1}{c}{$e [\%]$} & \multicolumn{1}{c}{$\bar{V}_x [m/s]$} & 
\multicolumn{1}{c}{$D [m^2/s]$} & \multicolumn{1}{c}{$e [\%]$} & \multicolumn{1}{c}{$\bar{V}_x 
[m/s]$} & \multicolumn{1}{c}{$D [m^2/s]$} & \multicolumn{1}{c}{$e [\%]$} \\ \hline
\textit{Ref. s.} & $7.49 \cdot 10^{-6}$ & $3.00 \cdot 10^{-6}$ & $-$ & $8.82 \cdot 10^{-6}$ & $5.29 
\cdot 10^{-6}$ & $-$ & $9.65 \cdot 10^{-6}$ & $7.72 \cdot 10^{-6}$ & $-$ & $1.03 \cdot 10^{-6}$ & 
$1.03 \cdot 10^{-6}$ & $-$ \\
\textit{M. 1} & $4.89 \cdot 10^{-7}$ & $6.27 \cdot 10^{-7}$ & $62.5$ & $6.02 \cdot 10^{-7}$ & $1.05 
\cdot 10^{-6}$ & $72.8$ & $7.03 \cdot 10^{-7}$ & $1.60 \cdot 10^{-6}$ & $79.5$ & $8.08 \cdot 
10^{-7}$ & $1.91 \cdot 10^{-6}$ & $80.3$ \\
\textit{M. 2} & $\mathit{1.52 \cdot 10^{-5}}$ & $\mathit{9.08 \cdot 10^{-7}}$ & $\mathit{44.2}$ & 
$\mathit{2.56 \cdot 10^{-5}}$ & $\mathit{1.58 \cdot 10^{-6}}$ & $\mathit{54.01}$ & $\mathit{3.79 
\cdot 10^{-5}}$ & $\mathit{8.79 \cdot 10^{-7}}$ & $\mathit{48.1}$ & $\mathit{4.02 \cdot 10^{-5}}$ & 
$\mathit{8.32 \cdot 10^{-7}}$ & $\mathit{42.9}$ \\
\textit{M. 3} & $4.51 \cdot 10^{-7}$ & $8.57 \cdot 10^{-7}$ & $87.6$ & $5.52 \cdot 10^{-7}$ & $1.46 
\cdot 10^{-6}$ & $108.8$ & $6.36 \cdot 10^{-7}$ & $2.34 \cdot 10^{-6}$ & $126.7$ & $7.24 \cdot 
10^{-7}$ & $2.94 \cdot 10^{-6}$ & $130.1$ \\
\textit{M. 4} & $\mathit{4.89 \cdot 10^{-7}}$ & $\mathit{6.27 \cdot 10^{-7}}$ & $\mathit{70.4}$ & 
$\mathit{6.02 \cdot 10^{-7}}$ & $\mathit{1.05 \cdot 10^{-6}}$ & $\mathit{59.2}$ & $\mathit{8.95 
\cdot 10^{-8}}$ & $\mathit{3.74 \cdot 10^{-5}}$ & $\mathit{49.2}$ & $\mathit{4.22 \cdot 10^{-5}}$ & 
$\mathit{3.53 \cdot 10^{-7}}$ & $\mathit{43.1}$ \\ \hline
\end{tabular}
\caption{Simulated ($Reference \ solution$) and estimated ($Methods \ 1-4$) values for average 
Darcy velocity $\bar{V}_x$, macrodispersion $D_{mac}$ and relative error $e$ values obtained for 
high permeability contrast simulations. $Method \ 1$ is the moments method, $Methods \ 2, \ 3$ and 
$4$ correspond to least squares method applied to the simulated PDF, CDF and the PDF peak data.}
\label{tab:VelDispErrHC}
\end{sidewaystable}
\restoregeometry

Summarizing the results of our numerical experiments, we can say that both correlation length and 
permeability contrast are triggering factors for non-Fickian transport behaviour. 
However, the impact of the correlation length on transport becomes more evident as the permeability 
contrast increases.

\subsection{Effect of Péclet} \label{sec:Peclet}
\begin{figure}[!htbp]
	\centering
	\includegraphics[width=0.45\linewidth]{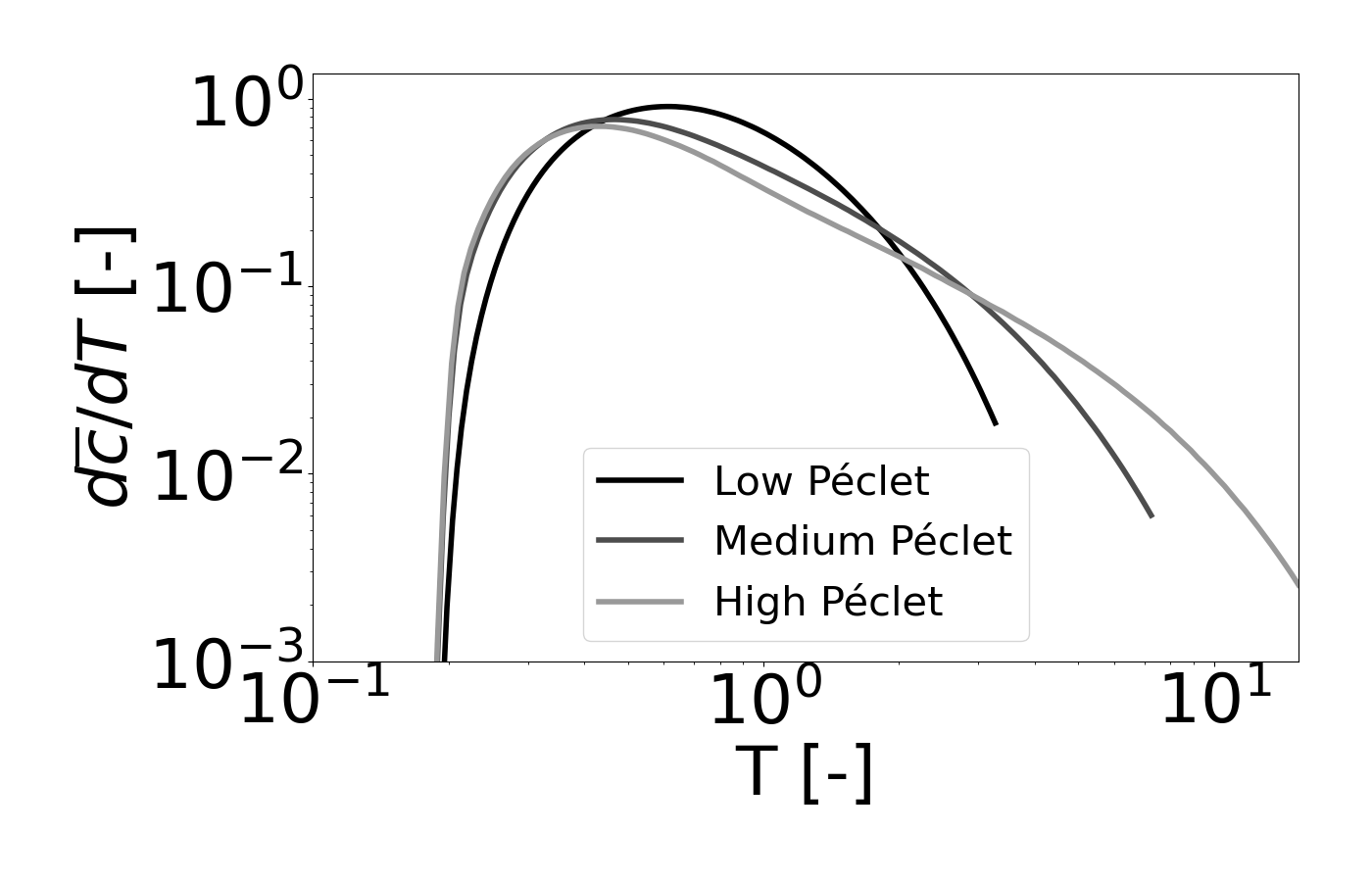}
	\includegraphics[width=0.45\linewidth]{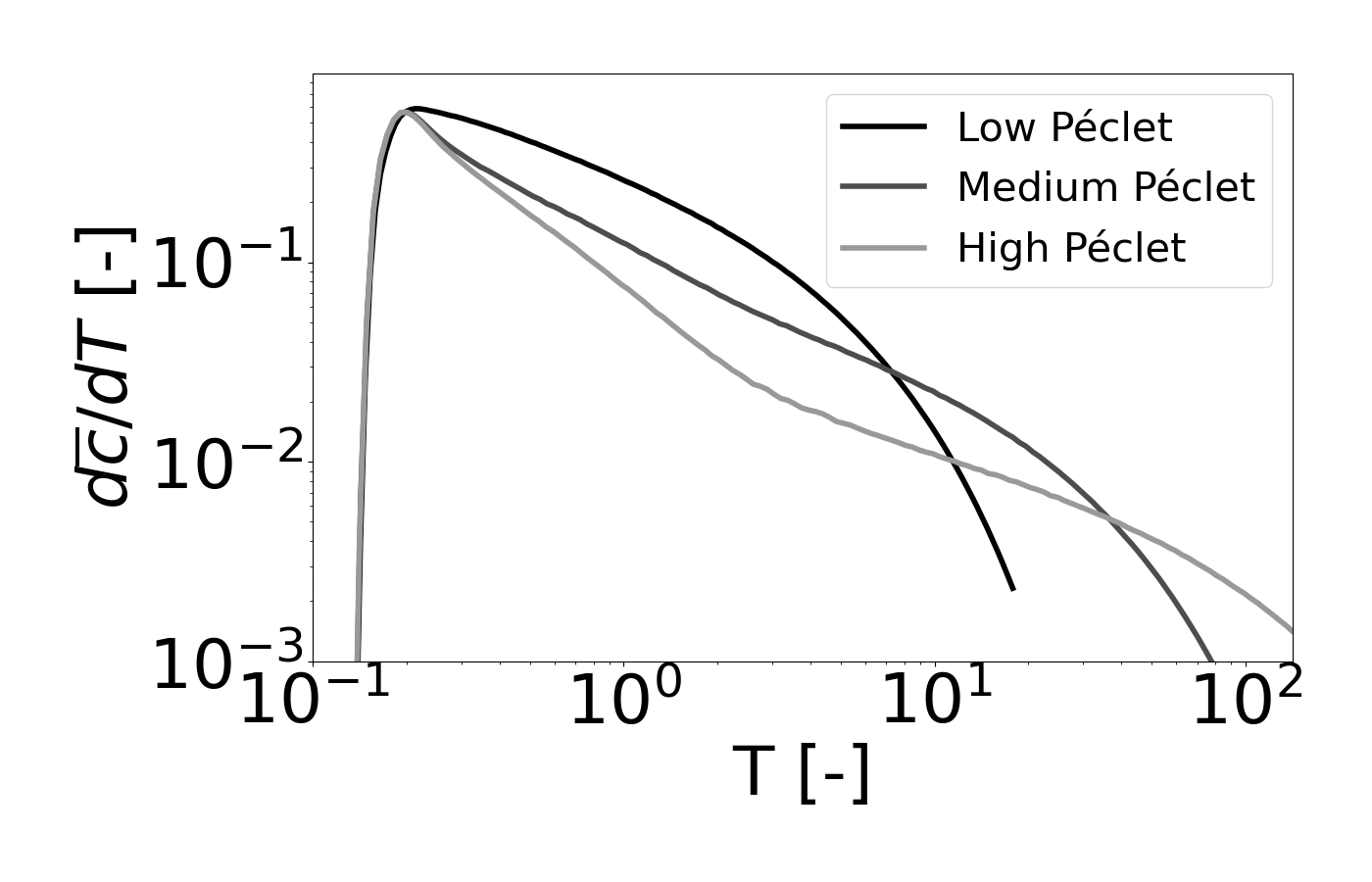}
	\caption{ Arrival times PDFs in low (left) and high (right) permeability contrast domains 
sharing the same geological structure ($\lambda_x=0.8$m) while characterised by a different Péclet. 
In case of low permeability contrast, the Péclet number ranges between $8 \cdot 10^2$ and $8 \cdot 
10^4$ while for high permeability contrast the Péclet ranges between $6 \cdot 10^3$ and $6 \cdot 
10^5$. Significant Péclet variation for these simulations was obtained by changing the molecular 
diffusion coefficient.}
	\label{fig:constVSvarMecDisp}
\end{figure}

We analyse here the effect of the Péclet number on our results. 
This means that not only the effect of different permeability fields (as in section 
\ref{sec:PermCont}) or correlation lengths (section \ref{sec:CL}) were tested, but also the effect 
of different solute composition is considered. 
This was performed by changing the molecular diffusion coefficient while keeping the 
average Darcy velocity and the correlation length $\lambda_x$ constant. Our tests were conducted by varying 
the Péclet ($Pe$) of each simulation by one order of magnitude. The variations in $Pe$ have a 
marked influence on the right tail of the arrival times distributions, as shown in 
\ref{fig:constVSvarMecDisp}. This result can be explained observing that diffusion effects 
become apparent in late arrivals, while early arrivals are conversely driven by advection-dominated 
processes. Left and right panels in figure \ref{fig:constVSvarMecDisp} show the combined effect of 
increasing Péclet (darker to lighter curves) in a low (left) and high (right) permeability contrast 
domain. The analysis of PDF curves for increasing Péclet yields comparable conclusions with the 
ones obtained in section \ref{sec:CL}: while the interplay between increasing Péclet numbers and 
the departure from Fickian behaviour is clear, the increase of permeability contrast still appears 
to play a predominant role.

\begin{figure}[!htpb]
	\centering
	\includegraphics[width=0.9\linewidth]{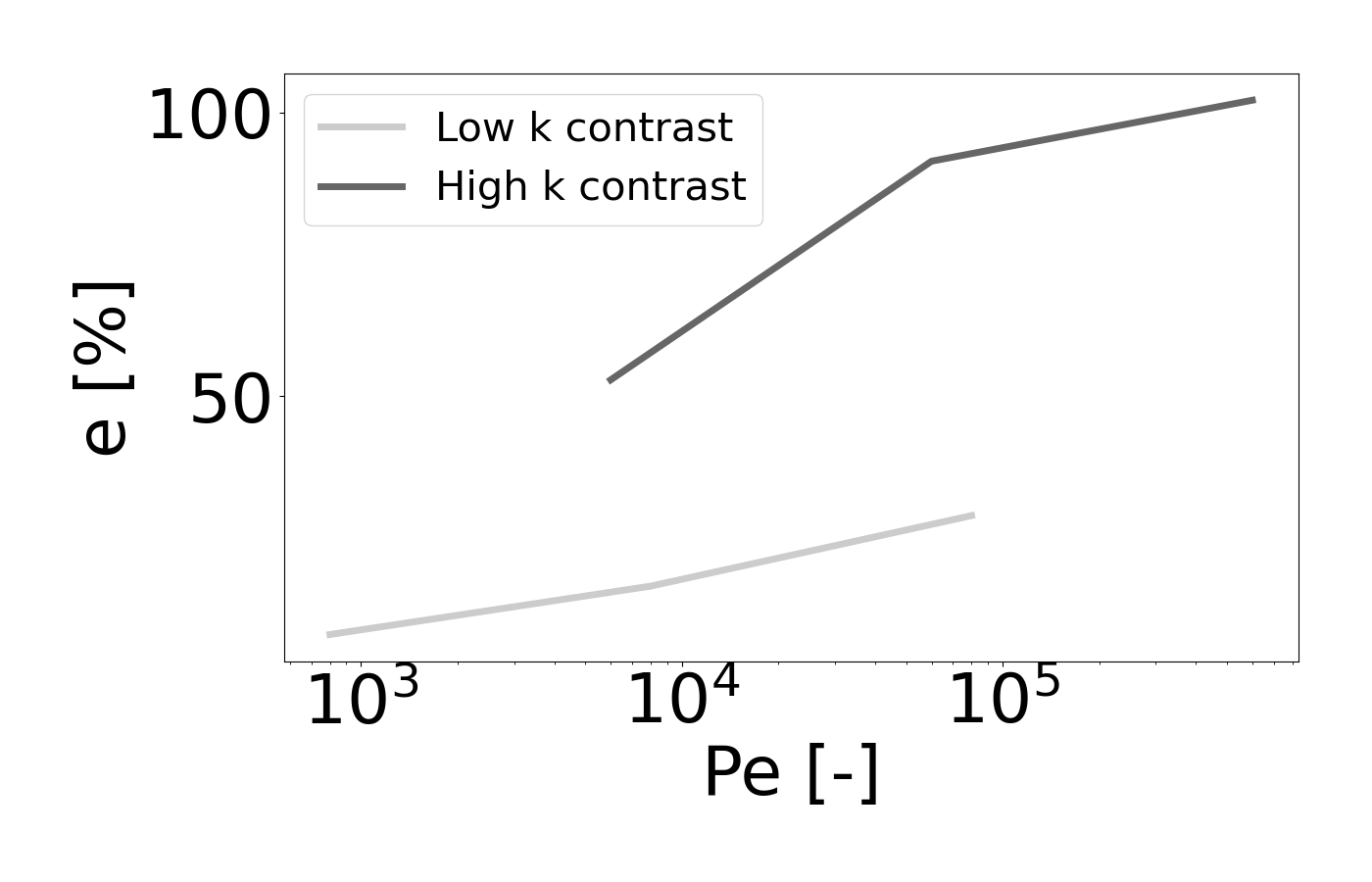}
	\caption{Péclet number vs relative error computed as in eq. \eqref{eq:RelErr}. The 
interplay between increasing Péclet numbers, permeability contrast and relative error is 
qualitatively similar to the one exhibited by the increasing longitudinal correlation length in 
fig. \ref{fig:LowHighKbtc}.}
	\label{fig:ErrPe}
\end{figure}

Figure \ref{fig:ErrPe} shows that the increase in permeability contrast by one order of magnitude 
exhibits a stronger control on transport behaviour than the increase in Péclet number by one 
magnitude order as for comparable $Pe$, the error associated with low permeability contrast 
simulations (light curve) is always lower than the error associated with the high permeability 
contrast (dark curve). The method considered for representing the ideal Fickian behaviour ($F(x)$ 
in eq. \eqref{eq:RelErr}) for the error curves is the statistical moments method (M.1 in tables 
\ref{tab:VelDispErrLC} and \ref{tab:VelDispErrHC}).

\clearpage

\section{Conclusions}
In this work we have explored the emergence of non-Fickian transport in random discontinuous 
permeability fields. These fields have a high degree of geological realism and present some relevant non-Fickian 
triggering features such as sharp contrast between facies and high connectivity degree. In 
particular, long channels enhance spatial velocity correlation and the larger and longer the 
correlated structures are, the more anomalous the transport will appear, with characteristics such 
as sharp concentration peak or early arrivals.

Flow and transport simulations have been performed over four sets of geological domains varying 
the assumed permeability contrast, correlation length and Péclet number. We compute the 
breakthrough curves as the flux-averaged concentration on the outlet section of the domain and 
analyse them as a key indicator of the transport behaviour. Statistics related to the breakthrough 
curves quantify the deviation from the advection dispersion model, with parameters calibrated with 
the method of moments and with global and local (near the peak) least-squares minimisation. We 
quantify the anomalous transport by comparing the mean square error and the resulting macroscopic 
parameters. This analysis led to the identification of sharp permeability contrast as the pivotal 
factor in creating fast flow channels where the solute arrival times tend to anticipate and 
overestimate the Fickian concentration peak. In fact, while Fickian transport is characterised by a 
single peak followed by an exponential-like tail, non-Fickian breakthrough curves present 
variations such as thinner peaks, followed longer power law tails. Our results show that the 
breakthrough curves trend can be subdivided into three stages: a steep peak of the early arrivals 
due to the channelling, followed by a power law decay due to the heterogeneous transport 
velocities, and eventually the exponential-like trend.  For moderate permeability contrasts,  
macrodispersion model can reasonably match only solute transport in low connected porous media. 
More advanced macro-models are needed to capture transport behaviour in highly connected porous 
media. The conclusions that can be drawn from the interpretation of these results are:
\begin{itemize}
	\item flow velocity in low permeability regions is homogeneously distributed around the 
corresponding peak values. In high permeability regions flow velocities display a left-skewed 
distribution, indicating the occurrence of low velocity regions in highly permeable media;
	\item the BTC variability observed between multiple realisations of the same geological 
setting is more evident at early times while it tends to disappear at late times; 
	\item our results combine error and uncertainty quantification metrics within the physical 
characterisation of the transport process. Emergence of non Fickian transport is quantified upon 
relying on relative error with respect to the prediction of a macrodispersive solution, where this latter can be 
obtained with diverse estimation strategies. Large relative errors and large confidence intervals 
for estimated parameters are indicative of the  unsuitability of the Inverse Gaussian distribution 
in interpreting the outcomes of high-resolution  numerical simulations, thus indicating non-Fickian 
features. 
	\item a Fickian macrodispersive model can match with reasonable accuracy solute arrival 
times in domains featuring conductivity values distributed over up to four orders of magnitudes. In 
such conditions the Fickian model provides a poor prediction of early arrivals, but can capture 
with good accuracy the peak and the late arrivals. Overall observed errors are in the order of 
10-20 $\%$. Lowest errors are associated with cases where the characteristic size associated with 
the medium heterogeneity is much smaller than the distance travelled by the solute.
	\item a hierarchy of non-Fickian triggering factors can be established: permeability 
contrast plays a primary role in determining the fate of the solute, while correlation length and 
Péclet number can be both considered secondary non-Fickian transport triggering factors.
\item while for Fickian or moderately non-Fickian transport the different  parameter estimation methods
(method of moments or least-squares-based methods) are equivalent, when a macrodispersion 
approximation is sought for significantly non-Fickian curves,
the choice of the fitting method is crucial as it can lead to very different effective
parameters and fitted curves. Although this is expected, due to the lack of validity of
the underlying model, it has important practical consequences for practitioners that
are nevertheless forced to use and fit macrodispersion effective parameters. Here,
the method of moments is built to preserve accurately the statistics but it could
significantly mispredict the early arrival peak as well as the long tails.

\end{itemize}

Future works will include the extension to more realistic injection scenarios,
variable density and hydrodynamic dispersion models, investigating the effect of different lithotype rules,
as well as interpreting the non-Fickian transport results with more sophisticated anomalous
transport models including spatial Markov processes \citep{sherman2021review} and Generalised
Multi-Rate Transfer equations \citep{municchi2020generalized}.

\section*{Declarations}

\subsection*{Conflict of Interests}
The authors have no relevant financial or non-financial interests to disclose.

\subsection*{Funding}
This work was supported by the Royal Society through the grant No. IES\textbackslash R3\textbackslash 170302.

\subsection*{Acknowledgements}
We would like to especially thank Prof. John Billingham for his insightful suggestions and for 
his detailed proofreading of this paper.

\clearpage

\bibliography{sn-bibliography}

\end{document}